\documentclass[preprint,showpacs,preprintnumbers,amsmath,amssymb]{revtex4}
\usepackage{epsfig}
\usepackage{graphicx}
\usepackage{dcolumn}
\usepackage{bm}
\usepackage{threeparttable}

\def\beq{\begin{equation}}
\def\eeq{\end{equation}}
\def\eeqn{\end{equation}}
\newcommand\iden{\leavevmode\hbox{\small1\normalsize\kern-.33em1}}

\newcommand{\bea} {\begin{eqnarray}}
\newcommand{\eea} {\end{eqnarray}}

\let\jnfont=\rm
\def\NPB#1,{{\jnfont Nucl.\ Phys.\ B }{\bf #1},}
\def\PLB#1,{{\jnfont Phys.\ Lett.\ B }{\bf #1},}
\def\EPJC#1,{{\jnfont Eur.\ Phys.\ Jour.\ C }{\bf #1},}
\def\PRD#1,{{\jnfont Phys.\ Rev.\ D }{\bf #1},}
\def\PRL#1,{{\jnfont Phys.\ Rev.\ Lett.\ }{\bf #1},}
\def\MPLA#1,{{\jnfont Mod.\ Phys.\ Lett.\ A }{\bf #1},}
\def\JPG#1,{{\jnfont J.\ Phys.\ G }{\bf #1},}
\def\CTP#1,{{\jnfont Commun.\ Theor.\ Phys.\ }{\bf #1},}
\def\JHEP#1,{{\jnfont JHEP \ }{\bf #1},}
\def\NPPS#1,{{\jnfont Nucl.\ Phys.\ Proc.\ Suppl.\ }{\bf #1},}
\def\CPC#1,{{\jnfont Comput.\ Phys.\ Commun.\ }{\bf #1},}
\def\CPL#1,{{\jnfont Chin.\ Phys.\ Lett. }{\bf #1},}
\def\APPB#1,{{\jnfont Acta\ Phys.\ Polon.\ B }{\bf #1},}

\def\lsim{\raise0.3ex\hbox{$<$\kern-0.75em\raise-1.1ex\hbox{$\sim$}}}
\def\gsim{\raise0.3ex\hbox{$>$\kern-0.75em\raise-1.1ex\hbox{$\sim$}}}
\def\PR#1,{{\jnfont Phys.\ Rept. }{\bf #1},}
\def\CHC#1,{{\jnfont Chin.\ Phys.\ C }{\bf #1},}

\begin{document}

\title{\ \\[10mm]$h\to\mu\tau$ and muon g-2 in the alignment limit of two-Higgs-doublet model}
\author{Lei Wang$^{1}$, Shuo Yang$^{2}$, Xiao-Fang Han$^{1}$}
 \affiliation{$^1$ Department of Physics, Yantai University, Yantai
264005, P. R. China\\
$^2$ Department of Physics, Dalian University, Dalian 116622, P. R.
China}


\begin{abstract}
We examine the $h\to \mu\tau$ and muon g-2 in the exact
alignment limit of two-Higgs-doublet model. In this case, the couplings of the SM-like Higgs
to the SM particles are the same as the Higgs couplings in the SM at the tree level,
and the tree-level lepton-flavor-violating coupling $h\mu\tau$ is
absent. We assume the lepton-flavor-violating $\mu\tau$ excess
observed by CMS to be respectively from the other neutral Higgses, $H$
and $A$, which almost degenerates with the SM-like Higgs at the 125
GeV. After imposing the relevant theoretical constraints and
experimental constraints from the precision electroweak data,
$B$-meson decays, $\tau$ decays and Higgs searches, we find that the muon
g-2 anomaly and $\mu\tau$ excess favor the small lepton Yukawa
coupling and top Yukawa coupling of the non-SM-like Higgs around
125 GeV, and the lepton-flavor-violating coupling is sensitive to
another heavy neutral Higgs mass. In addition, if the $\mu\tau$
excess is from $H$ around 125 GeV, the experimental data of the
heavy Higgs decaying into $\mu\tau$ favor $m_A>230$ GeV for a
relatively large $H\bar{t}t$ coupling.
\end{abstract}

\maketitle

\section{Introduction}
The ATLAS and CMS collaborations have probed the
lepton-flavor-violating (LFV) Higgs decay $h \to \mu\tau$ around 125
GeV at the LHC run-I \cite{150803372,160407730,cmstamu} and early
run-II \cite{161201644-5}. By the analysis of data sample
corresponding to an integrated luminosity of 20.3 fb$^{-1}$ at the
$\sqrt{s}$ = 8 TeV LHC, the ATLAS Collaboration found a mild
deviation of $1\sigma$ significance in the $h \to\mu\tau$ channel
and set an upper limit of $Br(h \to \mu\tau) < 1.43\%$ at 95\%
confidence level with a best fit $Br(h \to \mu \tau )$ = $(0.53 \pm
0.51)\%$ \cite{160407730}. Based
on the data sample corresponding to an integrated luminosity of 19.7
fb$^{-1}$ at the $\sqrt{s}$ = 8 TeV LHC, the CMS collaboration imposed an upper
limit of $Br(h\to \mu\tau ) < 1.51\%$ at 95\% confidence level, while the best fit value is
$Br(h\to \mu\tau)=(0.84^{+0.39}_{-0.37})\%$ with a small excess of
$2.4\sigma$ \cite{cmstamu}. At the $\sqrt{s}$ = 13 TeV LHC run-II
with an integrated luminosity of 2.3 fb$^{-1}$, the CMS
collaboration did not observe the excess and imposed an upper limit
of $Br(h\to \mu\tau ) < 1.2\%$ \cite{161201644-5}. However, the CMS
search result at the early LHC run-II can not definitely kill the
excess of $h\to \mu\tau$ due to the low integrated luminosity.

If the $h\to \mu\tau$ excess is not a statistical
fluctuation, the new physics with the LFV interactions can give a simple
explanation for the excess. On the other hand, the long-standing anomaly
of the muon anomalous magnetic moment (muon g-2) implies that the new
physics is connected to muons. The two excesses can be simultaneously
explained by the LFV Higgs interactions, such as the general
two-Higgs-doublet model (2HDM) with the LFV Higgs interactions.
There have been many studies on the $h\to\mu\tau$ excess in the
framework of 2HDM \cite{2hdmtamu1,2hdmtamu2,xp1601.02616} and some
other new physics models \cite{htumodel}.

In this paper, we discuss the excesses of $h\to \mu\tau$ and
muon g-2 in the exact alignment limit of the general 2HDM where 
one of the neutral Higgs mass eigenstates is aligned with the
direction of the scalar field vacuum expectation value (VEV)
\cite{alignment1}. In the interesting scenario, the SM-like Higgs
couplings to the SM particles are the same as the Higgs couplings in the SM at the tree level,
and the tree-level LFV coupling $h\mu\tau$ is absent. We assume the
$\mu\tau$ excess observed by CMS to be respectively from the
other neutral Higgses, $H$ and $A$, which almost degenerates with
the SM-like Higgs at the 125 GeV. In our discussions, we impose the
relevant theoretical constraints from the vacuum stability,
unitarity and perturbativity  as well as the experimental
constraints from the precision electroweak data, $B$-meson decays,
$\tau$ decays and Higgs searches.

Our work is organized as follows. In Sec. II we recapitulate the
alignment limit of 2HDM. In Sec. III we perform the numerical
calculations and discuss the muon g-2 anomaly and the $\mu\tau$
excess around 125 GeV after imposing the relevant theoretical
and experimental constraints. Finally, we give our conclusion in Sec.
IV.

\section{two-Higgs-doublet model and the alignment limit}
The alignment limit of 2HDM is defined as the limit in which one
of the two neutral CP-even Higgs mass eigenstates aligns with the
direction of the scalar field VEV \cite{alignment1}. The alignment
limit can be easily realized in the decoupling limit
\cite{1507.00933-12}, namely that all the non-SM-like Higgses
are very heavy. The possibility of alignment without decoupling
limit was first noted in \cite{1507.00933-12}, "re-invented" in
\cite{12074835,12100559,13052424} and further studied in
\cite{1507.00933-22,1507.00933-23,alignment1,aligndecp,alignment2}. The
alignment limit is basis-independent, and clearly exhibited in
the Higgs basis. The alignment limit also exists in the
Minimal Supersymmetric Standard Model which is a constrained incarnation of the general 2HDM. There are some
detailed discussions in \cite{13102248,14104969} and a very recent study in \cite{160800638}.

\subsection{Two-Higgs-doublet model in the Higgs basis}
The general Higgs potential is written as \cite{2h-poten}
\begin{eqnarray} \label{V2HDM} \mathrm{V} &=& \mu_{1}
(H_1^{\dagger} H_1) + \mu_{2} (H_2^{\dagger}
H_2) + \left[\mu_{3} (H_1^{\dagger} H_2 + \rm h.c.)\right]\nonumber \\
&&+ \frac{k_1}{2}  (H_1^{\dagger} H_1)^2 + \frac{k_2}{2}
(H_2^{\dagger} H_2)^2 + k_3 (H_1^{\dagger} H_1)(H_2^{\dagger} H_2) +
k_4 (H_1^{\dagger}
H_2)(H_2^{\dagger} H_1) \nonumber \\
&&+ \left[\frac{k_5}{2} (H_1^{\dagger} H_2)^2 + \rm h.c.\right]+
\left[k_6 (H_1^{\dagger} H_1)
(H_1^{\dagger} H_2) + \rm h.c.\right] \nonumber \\
&& + \left[k_7 (H_2^{\dagger} H_2) (H_1^{\dagger} H_2) + \rm
h.c.\right].
\end{eqnarray}
All $\mu_i$ and $k_i$ are real in the CP-conserving case. In the
Higgs basis,  the $H_1$ field has a VEV $v=$246 GeV, and the VEV of
$H_2$ field is zero. The two complex scalar doublets have the
hypercharge $Y = 1$,
\begin{equation}
H_1=\left(\begin{array}{c} G^+ \\
\frac{1}{\sqrt{2}}\,(v+\rho_1+iG_0)
\end{array}\right)\,, \ \ \
H_2=\left(\begin{array}{c} H^+ \\
\frac{1}{\sqrt{2}}\,(\rho_2+iA_0)
\end{array}\right).
\end{equation}

 The Nambu-Goldstone bosons $G^0$ and $G^+$ are eaten by the gauge bosons.
The $H^\pm$ and $A$ are the mass eigenstates of the charged Higgs
boson and CP-odd Higgs boson, and their masses are given by
\beq\label{mamhp} m_A^2=m^2_{H^{\pm}}+
\frac{1}{2}v^2(k_4-k_5). \eeq The physical CP-even Higgs
bosons $h$ and $H$ are the linear combinations of $\rho_1$ and
$\rho_2$,
\begin{equation}\label{mixalign}
\left(\begin{array}{c} \rho_1 \\
\rho_2
\end{array}\right)\, =\ \
\left(\begin{array}{c} ~s_\theta~~~~c_\theta \\
c_\theta~-s_\theta
\end{array}\right)\,
\left(\begin{array}{c} h \\
H
\end{array}\right),\,
\end{equation}
and their masses are given as \beq
m_{h,H}^2=\frac{1}{2}\left[m^2_{A} + (k_1 + k_5) v^2 \mp
\sqrt{[m^2_{A}+(k_5-k_1)v^2]^2+4k_6^2v^4}\right]. \eeq Where
$s_\theta\equiv\sin\theta$ and $c_\theta\equiv\cos\theta$, \beq
\cos\theta=\frac{-k_6v^2}{\sqrt{(m_H^2-m_h^2)(m_H^2-k_1v^2)}}.
\label{ctheta}\eeq

In this paper we take the light CP-even Higgs $h$ as the 125 GeV
Higgs. For $\cos\theta=0$, the mass eigenstates of CP-even Higgs
bosons are obtained from the Eq. (\ref{mixalign}), \beq
h=\rho_1,~~~~~H=-\rho_2, \eeq which is so called "alignment
limit". The Eq. (\ref{ctheta}) shows that the alignment limit can be
realized in two ways: $k_6=0$ or $m_H^2 \gg v^2$. The latter is
called the decoupling limit. In this paper we focus on the
former, which is the alignment without decoupling limit. In the
alignment limit, the $h$ couplings to gauge bosons are the same as the Higgs couplings in
the SM, and the $H$ has no couplings to gauge bosons.

\subsection{The Higgs couplings}
We can rotate the Higgs basis by a mixing
angle $\beta$,
\begin{equation}
\left(\begin{array}{c} \Phi_1 \\
\Phi_2
\end{array}\right)\, =\ \
\left(\begin{array}{c} ~c_\beta~~~-s_\beta \\
s_\beta~~~~~~c_\beta
\end{array}\right)\,
\left(\begin{array}{c} H_1 \\
H_2
\end{array}\right).\,
\end{equation}
Where $s_\beta\equiv\sin\beta$, $c_\beta\equiv\cos\beta$, and
$\tan\beta=v_2 /v_1$ with $v_2$ and $v_1$ being the VEVs of $\Phi_2$
and $\Phi_1$ and $v^2 = v^2_1 + v^2_2 = (246~\rm GeV)^2$.

The general Higgs potential is written as
\cite{2h-poten}
\begin{eqnarray} \label{V2HDM} \mathrm{V} &=& m_{11}^2
(\Phi_1^{\dagger} \Phi_1) + m_{22}^2 (\Phi_2^{\dagger}
\Phi_2) - \left[m_{12}^2 (\Phi_1^{\dagger} \Phi_2 + \rm h.c.)\right]\nonumber \\
&&+ \frac{\lambda_1}{2}  (\Phi_1^{\dagger} \Phi_1)^2 +
\frac{\lambda_2}{2} (\Phi_2^{\dagger} \Phi_2)^2 + \lambda_3
(\Phi_1^{\dagger} \Phi_1)(\Phi_2^{\dagger} \Phi_2) + \lambda_4
(\Phi_1^{\dagger}
\Phi_2)(\Phi_2^{\dagger} \Phi_1) \nonumber \\
&&+ \left[\frac{\lambda_5}{2} (\Phi_1^{\dagger} \Phi_2)^2 + \rm
h.c.\right]+ \left[\lambda_6 (\Phi_1^{\dagger} \Phi_1)
(\Phi_1^{\dagger} \Phi_2) + \rm h.c.\right] \nonumber \\
&& + \left[\lambda_7 (\Phi_2^{\dagger} \Phi_2) (\Phi_1^{\dagger}
\Phi_2) + \rm h.c.\right].
\end{eqnarray}
The parameters $m_{ij}$ and $\lambda_i$ are the linear combinations
of the parameters in the Higgs basis: $\mu_i$ and $k_i$. The
detailed expressions are introduced in \cite{alignment1,0504050}.
After spontaneous electroweak symmetry breaking, there are five
physical Higgses: two neutral CP-even $h$ and $H$, one neutral
pseudoscalar $A$, and two charged scalar $H^{\pm}$.

The general Yukawa interaction can be given as
 \bea
- {\cal L} &=& Y_{u1}\,\overline{Q}_L \, \tilde{{ \Phi}}_1 \,u_R
+\,Y_{d1}\,
\overline{Q}_L\,{\Phi}_1 \, d_R\, + \, Y_{\ell 1}\,\overline{L}_L \, {\Phi}_1\,e_R\nonumber\\
&&+\, Y_{u2}\,\overline{Q}_L \, \tilde{{ \Phi}}_2 \,u_R\,
+\,Y_{d2}\, \overline{Q}_L\,{\Phi}_2 \, d_R\,+\, Y_{\ell 2}
\overline{L}_L\, {\Phi}_2\,e_R \,+\, \mbox{h.c.}\,, \eea where
$Q_L^T=(u_L\,,d_L)$, $L_L^T=(\nu_L\,,l_L)$,
$\widetilde\Phi_{1,2}=i\tau_2 \Phi_{1,2}^*$, and $Y_{u1,2}$,
$Y_{d1,2}$ and $Y_{\ell 1,2}$ are $3 \times 3$ matrices in family
space.

To avoid the tree-level FCNC couplings of the quarks, we take \bea
&&Y_{u1}=c_u~\rho_u,~~Y_{u2}=s_u~ \rho_u, \nonumber\\
&&Y_{d1}=c_d~ \rho_d,~~Y_{d2}=s_d~ \rho_d, \eea where $c_u\equiv
\cos\theta_u$, $s_u\equiv \sin\theta_u$, $c_d\equiv \cos\theta_d$,
$s_d\equiv \sin\theta_d$ and $\rho_u$ ($\rho_d$) is the $3 \times 3$
matrix. For this choice, the interaction corresponds to the aligned
2HDM \cite{a2hm1,a2hm2}.

For the Yukawa coupling matrix of the lepton, we take \bea\label{klm}
&&X_{ii}=\frac{\sqrt{2}m_{\ell_i}}{v}(s_\beta+c_\beta
\kappa_\ell),\nonumber\\
&&X_{\tau\mu}=c_\beta \rho_{\tau\mu},\nonumber\\
&&X_{\mu\tau}=c_\beta \rho_{\mu\tau}.\eea Where $X=V_L Y_{\ell2}
V_R^{\dagger}$, and  $V_L$ $(V_R)$ is the unitary matrix which
transforms the interaction eigenstates to the mass eigenstates of
the left-handed (right-handed) lepton fields. The other nondiagonal
matrix elements of $X$ are zero.

The Yukawa couplings of the neutral Higgs bosons are given as
\bea\label{hffcoupling} &&
y_{hf_if_i}=\frac{m_{f_i}}{v}\left[\sin(\beta-\alpha)+\cos(\beta-\alpha)\kappa_f\right],\nonumber\\
&&y_{Hf_if_i}=\frac{m_{f_i}}{v}\left[\cos(\beta-\alpha)-\sin(\beta-\alpha)\kappa_f\right],\nonumber\\
&&y_{Af_if_i}=-i\frac{m_{f_i}}{v}\kappa_f~{\rm (for~u)},~~~~y_{Af_if_i}=i \frac{m_{f_i}}{v}\kappa_f~{\rm (for~d,~\ell)},\nonumber\\
&&y_{h\tau\mu}=\cos(\beta-\alpha)\frac{\rho_{\tau\mu}}{\sqrt{2}},~~~~~~~~y_{h\mu\tau}=\cos(\beta-\alpha)\frac{\rho_{\mu\tau}}{\sqrt{2}},\nonumber\\
&&y_{H\tau\mu}=-\sin(\beta-\alpha)\frac{\rho_{\tau\mu}}{\sqrt{2}},~~~~y_{H\mu\tau}=-\sin(\beta-\alpha)\frac{\rho_{\mu\tau}}{\sqrt{2}},\nonumber\\
&&y_{A\tau\mu}=i\frac{\rho_{\tau\mu}}{\sqrt{2}},~~~~~~~~~~~~~~~~~~~~y_{A\mu\tau}=i\frac{\rho_{\mu\tau}}{\sqrt{2}}.
\eea Where $\kappa_u\equiv-\tan(\beta-\theta_u)$,
 $\kappa_d\equiv-\tan(\beta-\theta_d)$. The $\kappa_\ell$ is a free input parameter, which is used to parameterize
the matrix element of the lepton Yukawa coupling, as shown in Eq. (\ref{klm}). In other words,
the matrix elements of the lepton Yukawa coupling are taken as the Eq. (\ref{klm}) in order to obtain
the Yukawa couplings of lepton in Eq. (\ref{hffcoupling}).

The neutral Higgs bosons couplings to the gauge bosons normalized to the
SM Higgs boson are given by \beq y^h_{V}=\sin(\beta-\alpha),~~~
y^H_{V}=\cos(\beta-\alpha),\label{hvvcoupling}\eeq where $V$ denotes
$Z$ and $W$.

In the exact alignment limit, namely $\cos(\beta-\alpha)=0$, the Eq.
(\ref{hffcoupling}) and Eq. (\ref{hvvcoupling}) show that the 125
GeV Higgs ($h$) has the same couplings to the fermions and gauge
bosons as the SM values, and the tree-level LFV couplings are absent. The
heavy CP-even Higgs ($H$) has no coupling to the gauge bosons, and
there are the tree-level LFV couplings for the $A$ and $H$.

\section{Numerical calculations and discussions}
\subsection{Numerical calculations}
In the exact alignment limit, the SM-like Higgs has no tree-level
LFV coupling. In order to explain the $h\to \mu\tau$ excess
reported by CMS, we assume the signal to be respectively from $H$ and
$A$, which almost degenerates with the SM-like Higgs at the 125 GeV.
Here we take two scenarios simply: (i) $m_A$=126 GeV and (ii)
$m_H$=126 GeV.

In our calculations, the other involved parameters are randomly
scanned in the following ranges:
\begin{eqnarray}
&&-(400~{\rm GeV})^2 \leq m_{12}^2 \leq (400~{\rm GeV})^2,~~~0.1\leq\tan\beta\leq 10,\nonumber\\
&&100 {\rm\  GeV} \leq ~m_{H^\pm}  \leq 700  {\rm\  GeV},\nonumber\\
&&0\leq \kappa_u \leq 1.2, ~~~ -150\leq \kappa_\ell \leq 150,~~~-0.3\leq\rho_{\tau\mu}\leq 0.3\nonumber\\
&&{\rm Scenario~i}:~~ m_A=126~{\rm GeV}, ~~150 {\rm\  GeV} \leq ~m_{H}  \leq 700 {\rm\  GeV},~~ \rho_{\mu\tau}=-\rho_{\tau\mu},\nonumber\\
&&{\rm Scenario~ii}:~m_H=126~{\rm GeV},~~150 {\rm\  GeV} \leq ~m_{A}
\leq 700 {\rm\  GeV},~~ \rho_{\mu\tau}=\rho_{\tau\mu}.
\end{eqnarray}

In order to relax the constraints from the observables of down-type
quarks, we take $\kappa_d=0$. For the cases of $m_A=126$ GeV and
$m_H$=126 GeV, we respectively take $\rho_{\mu\tau}=-\rho_{\tau\mu}$
and $\rho_{\mu\tau}=\rho_{\tau\mu}$ to produce a large positive
contribution to the muon g-2. The pseudoscalar $A$ can give the positive
contributions to the muon g-2 via the two-loop Barr-Zee diagrams
with the lepton-flavor-conserving (LFC) coupling. Therefore, we take
$\mid\kappa_\ell\mid<150$ to examine the possibility of explaining
the muon g-2. In the exact alignment limit, the $h\tau\bar{\tau}$
coupling is independent on $\kappa_\ell$ and equals the SM value.
However, the $A\tau\bar{\tau}$ and $H\tau\bar{\tau}$ couplings can
reach 1.08 and be slightly larger than 1 for
$\mid\kappa_\ell\mid=150$, which can not lead to the problem on the
perturbativity due to the suppression of the loop factor. In
addition, for such large $\kappa_\ell$ the $Br(A\to \tau\bar{\tau})$
and $Br(H\to \tau\bar{\tau})$ can reach 1. Due to $\kappa_d=0$ and
$\cos(\beta-\alpha)=0$, the cross sections of $A$ and $H$ are equal to
zero in the $b\bar{b}$ associated production mode and vector boson fusion
production mode. However, the searches for $gg\to A/H \to \tau\bar{\tau}$ can
give the constraints on $\kappa_u$. We will discuss the constraints in the following item
(5).

During the scan, we consider the following experimental constraints
and observables:

{\bf (1) Theoretical constraints and precision electroweak data}. We
use $\textsf{2HDMC-1.6.5}$ \cite{2hc-1} to implement the theoretical
constraints from the vacuum stability, unitarity and
coupling-constant perturbativity, as well as the constraints from
the oblique parameters ($S$, $T$, $U$) and $\delta\rho$.

{\bf (2) $B$-meson decays and $R_b$}. Although the tree-level FCNCs
in the quark sector are absent, they will appear at the one-loop
level in this model. We consider the constraints of $B$-meson decays
from $\Delta m_{B_s}$, $\Delta m_{B_d}$, $B\to X_s\gamma$, and
$B_s\to \mu^+\mu^-$, which are respectively calculated using the
formulas in \cite{deltmq,bsr,bsmumu}. In addition, we consider the $R_b$
constraints, which is
calculated following the formulas in \cite{rb}. In fact, in this
paper we take $\kappa_d=0$ and $0\leq\kappa_u\leq 1.2$, which will
relax the constraints from the bottom-quark observables sizably.

{\bf (3) $\tau$ decays}. In this model, the non-SM-like Higgses have
the tree-level LFV couplings to $\tau$ lepton, and the LFC
couplings to lepton can be sizably enhanced for
$-150\leq\kappa_\ell\leq150$. Therefore, some $\tau$ decay processes
can give very strong constraints on the model.

\begin{itemize}
\item[(i)] $\tau\to 3\mu$. In the exact alignment limit, the LFV $A\tau\mu$ and $H\tau\mu$ couplings
generate the $\tau\to\mu^+\mu^-\mu^-$ process at the tree level, and the corresponding Feynman diagrams
are shown in Fig. \ref{fmtta3mu}.
The branching ratio of $\tau\to 3\mu$ is given as \cite{ta3mu} \beq
\frac{Br(\tau\to3\mu)}{Br(\tau\to\mu\bar{\nu}\nu)}=\sum_{\phi_1,\phi_2=A,H}\frac{I(\phi_1,\phi_2)}{64G_F^2},\eeq
where \bea
I(\phi_1,\phi_2)=&&2\frac{y_{\phi_1\mu\tau}y_{\phi_1\mu\mu}^*}{m_{\phi_1}^2}
\frac{y_{\phi_2\mu\tau}^*y_{\phi_2\mu\mu}}{m_{\phi_2}^2}+
2\frac{y_{\phi_1\tau\mu}y_{\phi_1\mu\mu}^*}{m_{\phi_1}^2}
\frac{y_{\phi_2\tau\mu}^*y_{\phi_2\mu\mu}}{m_{\phi_2}^2}\nonumber\\
&&+\frac{y_{\phi_1\mu\tau}y_{\phi_1\mu\mu}}{m_{\phi_1}^2}
\frac{y_{\phi_2\mu\tau}^*y_{\phi_2\mu\mu}^*}{m_{\phi_2}^2}+
\frac{y_{\phi_1\tau\mu}y_{\phi_1\mu\mu}}{m_{\phi_1}^2}
\frac{y_{\phi_2\tau\mu}^*y_{\phi_2\mu\mu}^*}{m_{\phi_2}^2}.\label{ta3mu}\eea
The current experimental upper bound of $Br(\tau\to 3\mu)$ is
\cite{expta3mu}, \beq Br(\tau\to 3\mu) < 2.1\times 10^{-8}. \eeq
From the Eq. (\ref{ta3mu}), we can find that the experimental data of $Br(\tau\to
3\mu)$ will give the very strong constraints on the products,
$\rho_{\tau\mu}\times \kappa_\ell$ and $\rho_{\mu\tau}\times
\kappa_\ell$.


\begin{figure}[tb]

 \epsfig{file=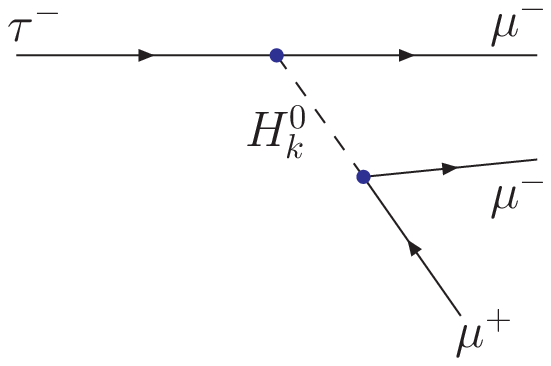,height=3.5cm}
\vspace{-0.5cm} \caption{The main Feynman diagrams of $\tau^-\to\mu^-\mu^-\mu^+$.
The two $\mu^-$ in final states can be exchanged.
In the exact alignment limit $H^0_k$ denotes $A$ and $H$.}
 \label{fmtta3mu}
\end{figure}

\begin{figure}[tb]

 \epsfig{file=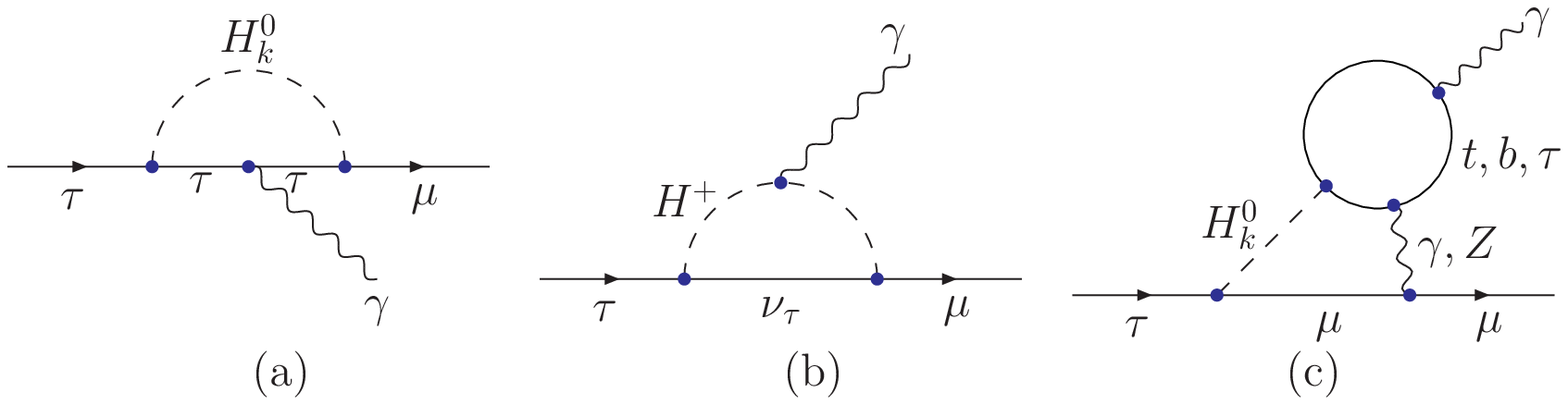,height=3.5cm}
\vspace{-0.5cm} \caption{The main Feynman diagrams of $\tau\to\mu\gamma$.
In the exact alignment limit $H^0_k$ denotes $A$ and $H$.}
 \label{fmttamur}
\end{figure}

\item[(ii)] $\tau\to \mu\gamma$. The main Feynman diagrams of $\tau\to \mu\gamma$ in the model
are shown in Fig. \ref{fmttamur}. In the exact alignment limit, the
SM-like Higgs has no tree-level LFV coupling, and the heavy CP-even
Higgs couplings to the gauge bosons are equal to zero. Therefore, the
SM-like Higgs does not contribute to the $\tau\to
\mu\gamma$, and the $\tau\to \mu\gamma$ can not be corrected
via the two-loop Barr-Zee diagrams with the $W$ loop. The $Br(\tau\to \mu\gamma)$ in this model is given by \beq
\frac{{\rm BR}(\tau \rightarrow \mu \gamma)}{{\rm BR}(\tau
\rightarrow \mu \bar{\nu} \nu)}
  =\frac{48\pi^3\alpha\left(|A_{1L0}+A_{1Lc}+A_{2L}|^2+|A_{1R0}+A_{1Rc}+A_{2R}|^2\right)}
  {G_F^2},
\eeq where $A_{1L0}$, $A_{1Lc}$, $A_{1R0}$ and $A_{1Rc}$ are from
the one-loop diagrams with the Higgs boson and $\tau$ lepton
\cite{2hdmtamu2}, \bea\label{tamura1}
A_{1L0}&=&\sum_{\phi=H,~A}\frac{y^{*}_{\phi\;\tau \mu}}{16\pi^2
m_\phi^2}
  \left[y^{*}_{\phi\;\tau\tau}\left(\log\frac{m_\phi^2}{m_\tau^2}-\frac{3}{2}\right)
    +\frac{y_{\phi\;\tau\tau}}{6}\right], \\
A_{1Lc}&=&-\frac{(\rho^{e\dagger} \rho^e)^{\mu \tau}}{192\pi^2
m_{H^-}^2}, \\\label{tamura1r} A_{1R0}&=&A_{1L0}\left({y^{*}_{\phi\;
\tau\mu}\rightarrow y_{\phi\; \mu\tau},
    ~~y_{\phi \;\tau \tau} \leftrightarrow y^{*}_{\phi\; \tau\tau}}\right),\\
A_{1Rc}&=&0. \eea The $A_{2L}$ and $A_{2R}$ are from the two-loop
Barr-Zee diagrams with the third-generation fermion loop
\cite{2hdmtamu2}, \bea\label{tamura2}
  A_{2L}&&=-\sum_{\phi=H,A;f=t,b,\tau}
  \frac{N_C Q_f \alpha}{8\pi^3}
  \frac{y^{*}_{\phi\;\tau\mu}}{m_\tau m_{f}}
  \left[
  Q_f\left\{
    {\rm Re} (y_{\phi\; ff})
    F_H\left(x_{f\phi}\right)
  - i {\rm Im} (y_{\phi\; ff})
  F_A\left(x_{f\phi}\right)\right\}\right.
  \nonumber \\
  &&\left. +\frac{(1-4 s_W^2)(2T_{3f}-4Q_f s_W^2)}{16s_W^2 c_W^2}
  \left\{
    {\rm Re} (y_{\phi\; ff})
    \tilde{F}_H\left(x_{f\phi},x_{fZ}\right)
  - i {\rm Im} (y_{\phi\; ff})
  \tilde{F}_A\left(x_{f\phi},x_{f Z}\right)\right\}\right], \nonumber \\
  A_{2R} &=&A_{2L}\left(y^{*}_{\phi\;\tau\mu}\rightarrow
  y_{\phi\;\mu\tau},~i\rightarrow -i\right),
\label{Barr-Zee}\eea where $T_{3f}$ denotes the isospin of the
fermion,  and
\begin{align}
F_{H}(y)&=\frac{y}{2}\int_0^1 dx \frac{1-2x(1-x)}{x(1-x)-y}\log
\frac{x(1-x)}{y}
~~({\rm for}~\phi=H), \nonumber\\
F_A(y) &=\frac{y}{2}\int_0^1 dx \frac{1}{x(1-x)-y}\log
\frac{x(1-x)}{y}~~
({\rm for}~\phi=A), \nonumber\\
  \tilde{F}_H(x,y)&=\frac{xF_H(y)-yF_H(x)}{x-y},\nonumber\\
  \tilde{F}_A(x,y)&=\frac{xF_A(y)-yF_A(x)}{x-y}.
\end{align}
The two terms of $A_{2L}$ come from the effective $\phi
\gamma\gamma$ vertex and $\phi Z\gamma$ vertex induced by the
third-generation fermion loop. The current experimental data give an
upper bound of $Br(\tau\to\mu\gamma)$ \cite{exptamur}, \beq
Br(\tau\to\mu\gamma) < 4.4\times 10^{-8}. \eeq

\begin{figure}[tb]

 \epsfig{file=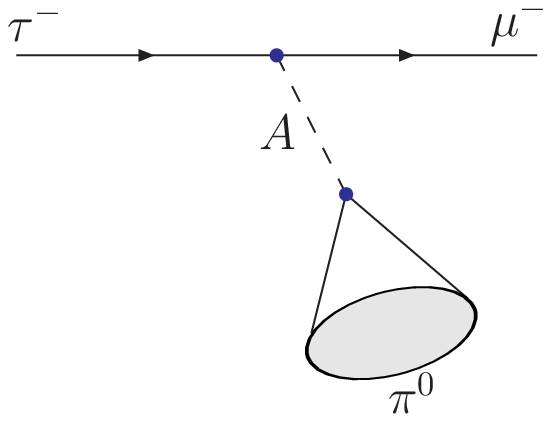,height=3.5cm}
\vspace{-0.2cm} \caption{The main Feynman diagram of $\tau\to\mu\pi^0$.}
 \label{fmttamupi}
\end{figure}
\item[(iii)] $\tau\to \mu\pi^0$. The $\tau$ can decay into a lepton and a pseudoscalar meson at the tree level
via the CP-odd Higgs with the LFV couplings, such as $\tau\to \mu\pi^0$.
The corresponding Feynman diagrams are shown in Fig.
\ref{fmttamupi}. The width of $\tau\to \mu\pi^0$ is given as
\cite{tamupi}, \beq\Gamma(\tau\to\mu\pi^0)=\frac{f^2_\pi m^4_\pi
m_\tau}{512\pi m^4_A
v^2}(|\rho_{\tau\mu}|^2+|\rho_{\mu\tau}|^2)(\kappa_u+\kappa_d)^2.
\eeq The current upper bound of $Br(\tau\to\mu\pi^0)$ is
\cite{exptamupi}, \beq Br(\tau\to\mu\pi^0) < 1.1\times 10^{-7}. \eeq

\end{itemize}

{\bf(4) muon g-2}. The dominant contributions to the muon g-2 are
from the one-loop diagrams
 with the Higgs LFV coupling \cite{mua1loop}, and the corresponding Feynman diagrams
can be obtained by replacing the initial states $\tau$ with $\mu$ in Fig. \ref{fmttamur} (a) and
Fig. \ref{fmttamur} (b). In the exact alignment
 limit,
\bea
  \delta a_{\mu1}^{LFV}&=&\frac{m_\mu m_\tau \rho_{\mu\tau}\rho_{\tau\mu}}{16\pi^2}
  \left[\frac{(\log\frac{m_H^2}{m_\tau^2}-\frac{3}{2})}{m_H^2}
  -\frac{\log(\frac{m_A^2}{m_\tau^2}-\frac{3}{2})}{m_A^2}
\right].
  \label{mua1}
\eea

At the one-loop level, the diagrams with the Higgs LFC coupling can
also give the contributions to the muon g-2, especially for a large
lepton Yukawa coupling \cite{mua1looplfc}. The corresponding
Feynman diagrams can be obtained by replacing $\tau$ in the initial
state and loop with $\mu$ in Fig. \ref{fmttamur} (a) as well as
replacing the initial state $\tau$ with $\mu$ and $\nu_\tau$ in the
loop with $\nu_\mu$ in Fig. \ref{fmttamur} (b). The contributions
from the one-loop diagrams with the Higgs LFC coupling are given as
\beq
    \Delta a_{\mu1}^{LFC} =
    \frac{1}{8 \pi^2 } \, \sum_{\phi = h,~ H,~ A ,~ H^\pm}
    |y_{\phi\mu\mu}|^2  r_{\phi\mu} \, f_\phi(r_{\phi\mu}),
\label{amuoneloop}
\end{equation}
where $r_{\phi\mu} =  m_\mu^2/m_\phi^2$ and $y_{H^\pm\mu\mu}=
y_{A\mu\mu}$. For $r_{\phi\mu}\ll$ 1, \beq
    f_{h,H}(r) \simeq- \ln r - 7/6,~~
    f_A (r) \simeq \ln r +11/6, ~~
    f_{H^\pm} (r) \simeq -1/6.
    \label{oneloopintegralsapprox3}
\eeq

The muon g-2 can be corrected by the two-loop Barr-Zee diagrams with
the fermions loops by replacing the initial state $\tau$ with $\mu$
in Fig. \ref{fmttamur} (c). Further replacing the fermion loop with
$W$ loop, we obtain the two-loop Barr-Zee diagrams with $W$ loop
which can contribute to muon g-2 for the SM-like Higgs $h$ as the
mediator in the exact alignment limit. Using the well-known
classical formulates in \cite{mua2loop}, the main contributions of two-loop Barr-Zee diagrams
 in the exact alignment limit are given as
\bea \label{mua2} \delta a_{\mu2} &=&-\frac{\alpha
m_\mu}{4\pi^3m_f}\sum_{\phi=h,H,A;f=t,b,\tau} N_f^c~Q_f^2~
y_{\phi\mu\mu}~ y_{\phi ff}~ F_\phi(x_{f\phi})
\nonumber\\
&&+\frac{\alpha m_\mu}{8\pi^3v}\sum_{\phi=h} y_{\phi\mu\mu}~g_{\phi
WW} \left[3F_H\left(x_{W\phi}\right)
    +\frac{23}{4} F_A\left(x_{W\phi}\right)
     \right.\nonumber\\
    &&\left.
+\frac{3}{4} G\left(x_{W\phi}\right) +\frac{m_\phi^2}{2
m_W^2}\left\{
    F_H\left(x_{W\phi}\right)-F_A\left(x_{W\phi}\right)
    \right\}\right],
\eea where $x_{f\phi}=m_{f}^2/m_\phi^2$, $x_{W\phi}=m_W^2/m_\phi^2$,
$g_{h WW}=1$ and \beq
  G(y)=-\frac{y}{2}\int_0^1 dx \frac{1}{x(1-x)-y}\left[
    1-\frac{y}{x(1-x)-y}\log \frac{x(1-x)}{y}
    \right].
\eeq
 The experimental value of muon g-2 excess is \cite{muaexp}
\beq \delta a_{\mu} =(26.2\pm8.5) \times 10^{-10}. \eeq

{\bf (5) Higgs searches experiments}.
\begin{itemize}
\item[(i)] Non-observation of additional Higgs bosons. We employ
$\textsf{HiggsBounds-4.3.1}$ \cite{hb} to implement the exclusion
constraints from the neutral and charged Higgses searches at LEP,
Tevatron and LHC at 95\% confidence level.

\item[(ii)] The global fit to the 125 GeV Higgs signal data. In the exact alignment
limit, the SM-like Higgs has the same coupling to the gauge boson
and fermions as the Higgs couplings in the SM, which is favored by the 125 GeV Higgs signal
data. However, in order to explain the $\mu\tau$ excess around 125
GeV, we assume that the $A$ ($H$) almost degenerates with the SM-like
Higgs at the 125 GeV. Since the mass splitting of $A$ ($H$) and $h$
is smaller than the mass resolution of detector, $A$ ($H$) can
affect the global fit to the 125 GeV Higgs signal data. Following
the method in \cite{chi}, we perform a global fit to the 125 GeV
Higgs data of 29 channels, which are given in the appendix A. The signal strength for
a channel is defined as \beq
\mu_i=\sum_{\hat{H}=h,~\phi}\epsilon_{gg\hat{H}}^i
R_{gg\hat{H}}+\epsilon_{VBF\hat{H}}^i
R_{VBF\hat{H}}+\epsilon_{V\hat{H}}^i
R_{V\hat{H}}+\epsilon_{t\bar{t}\hat{H}}^i R_{t\bar{t}\hat{H}}. \eeq
 Where $R_{j}=(\sigma \times BR)_j/(\sigma\times BR)_j^{SM}$
with $j$ denoting the partonic process
$gg\hat{H},~VBF\hat{H},~V\hat{H},$ or $t\bar{t}\hat{H}$.
$\epsilon_{j}^i$ denotes the assumed signal composition of the
partonic process $j$. If $A$ ($H$) almost degenerates with the SM-like Higgs, $\phi$ denotes
$A$ ($H$). For an uncorrelated observable $i$, \beq
\chi^2_i=\frac{(\mu_i-\mu^{exp}_i)^2}{\sigma_i^2}, \eeq where
$\mu^{exp}_i$ and $\sigma_i$ denote the experimental central value
and uncertainty for the $i$-channel. We retain the uncertainty
asymmetry in the calculation. For the two correlated observables, we
take \beq \chi^2_{i,j}=\frac{1}{1-\rho^2}
\left[\frac{(\mu_i-\mu^{exp}_i)^2}{\sigma_i^2}+\frac{(\mu_j-\mu^{exp}_j)^2}{\sigma_j^2}
-2\rho\frac{(\mu_i-\mu^{exp}_i)}{\sigma_i}\frac{(\mu_j-\mu^{exp}_j)}{\sigma_j}\right],
\eeq where $\rho$ is the correlation coefficient. We sum over
$\chi^2$ in the 29 channels, and pay particular attention to the
surviving samples with $\chi^2-\chi^2_{\rm min} \leq 6.18$, where
$\chi^2_{\rm min}$ denotes the minimum of $\chi^2$. These samples
correspond to the 95.4\% confidence level region in any
two-dimension plane of the model parameters when explaining the
Higgs data (corresponding to the $2\sigma$ range).

\item[(iii)] The Higgs decays into $\tau\mu$. In the exact
alignment limit, the $\mu\tau$ excess around 125 GeV is from
$A~(H)\to \tau\mu$ where $A$ ($H$) almost degenerates with the
SM-like Higgs. The width of $A~(H)\to \mu\tau$ is given by \beq
\Gamma(A~(H)\to \mu\tau)=\frac{
(\rho_{\mu\tau}^2+\rho_{\tau\mu}^2)m_{A~(H)}}{16\pi}. \eeq
We take the best fit value of $Br(h\to \mu\tau)=(0.84^{+0.39}_{-0.37})\%$ 
based on the CMS search for the $h\to \mu\tau$
at the LHC run-I.
Since the $\mu\tau$ excess is assumed to be from the $A$ ($H$), we require the
production rates of $pp\to A~(H)\to \mu\tau$ to vary from
$\sigma(pp\to h)\times 0.1\%$ to $\sigma(pp\to h)\times 1.62\%$.

In addition, the CMS collaboration did not publish the bound on the
heavy Higgs decaying into $\mu\tau$. Ref. \cite{xp1601.02616} gave the
bound on the production rate of $pp\to\phi\to\mu\tau$ by recasting
results from the original $h\to\mu\tau$ analysis of CMS.

\end{itemize}

\subsection{Results and discussions}
\begin{figure}[tb]
 \epsfig{file=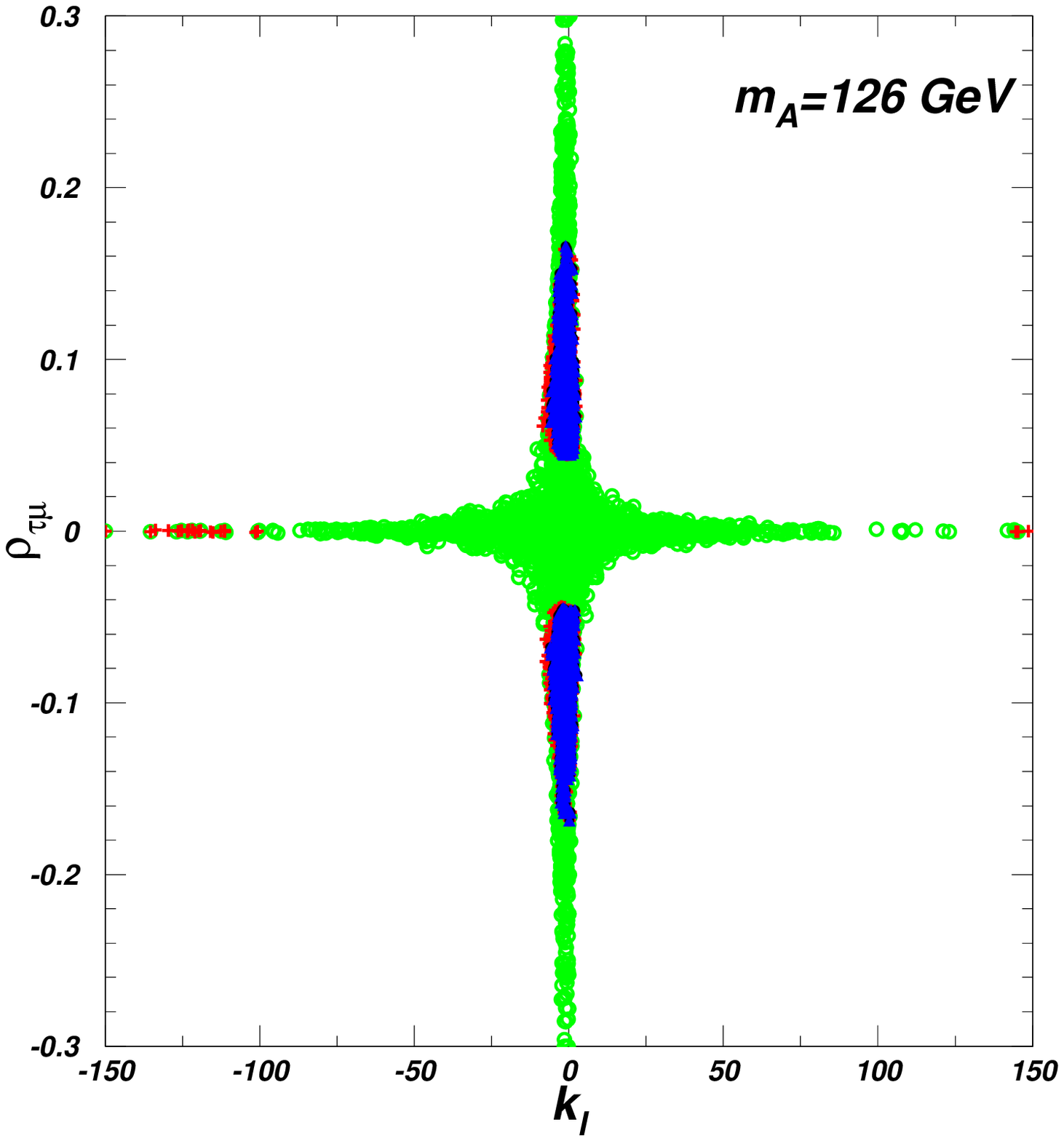,height=7.0cm}
  \epsfig{file=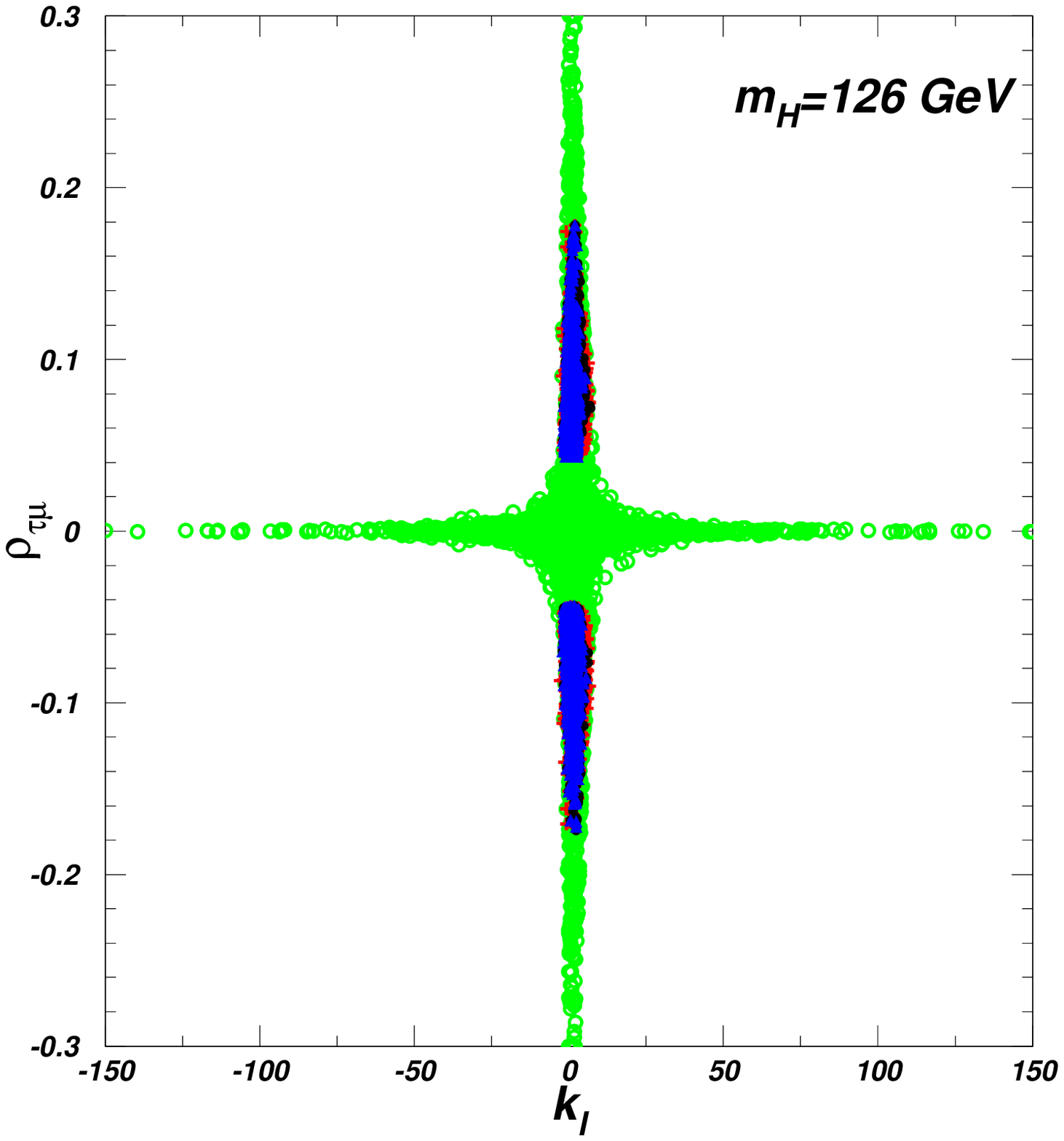,height=7.0cm}
 \epsfig{file=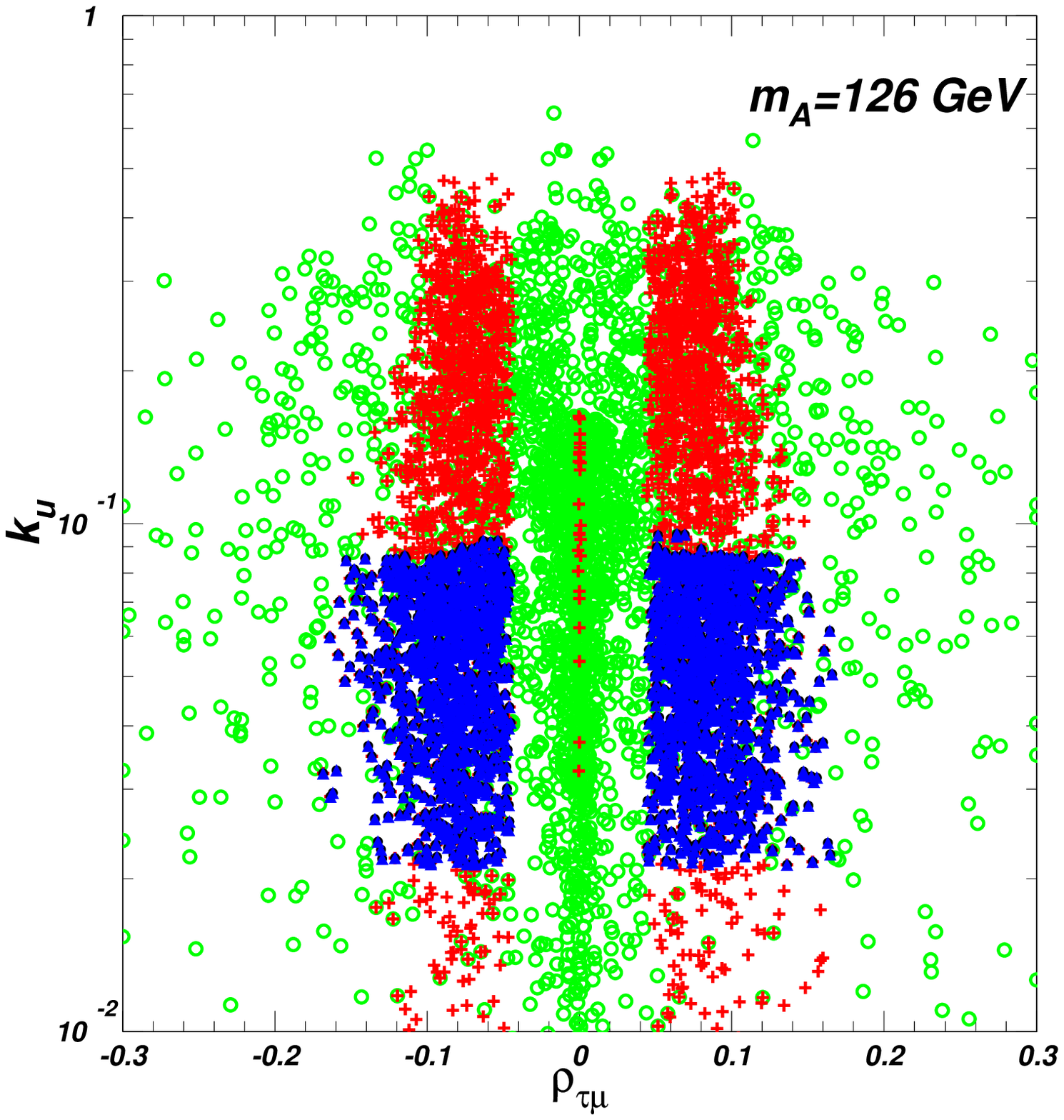,height=7.1cm}
  \epsfig{file=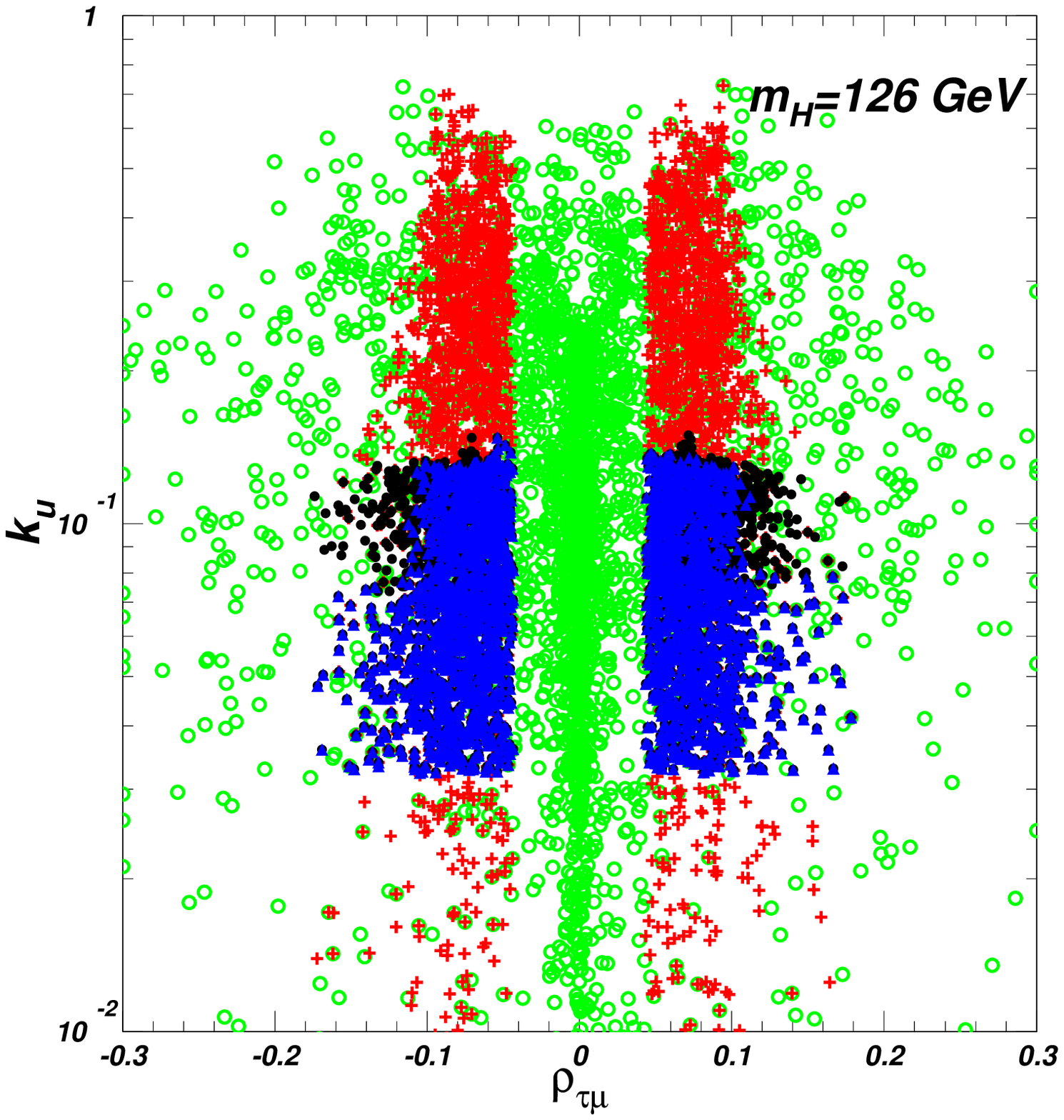,height=7.0cm}
\vspace{-0.5cm} \caption{The surviving samples projected on the
planes of $\rho_{\tau\mu}$ versus $\kappa_\ell$ and $\kappa_u$
versus $\rho_{\tau\mu}$. The circles (green) are allowed by the
"pre-muon g-2" constraints: theoretical constraints, precision
electroweak data, $R_b$, $B$ meson decays, $\tau$ decays, the
exclusion limits of Higgses, and the 125 GeV Higgs data; the pluses
(red) allowed by the pre-muon g-2 and muon g-2 excess; the bullets
(black) and triangles (blue) allowed by the pre-muon g-2, the
 muon g-2 anomaly and $\mu\tau$ excess around 125 GeV, and the
 triangles (blue)
further allowed by the experimental constraints of the heavy Higgs
decaying into $\mu\tau$.} \label{cpvscp}
\end{figure}

In Fig. \ref{cpvscp}, we project the surviving samples on the planes
of $\rho_{\tau\mu}$ versus $\kappa_\ell$ and $\kappa_u$ versus
$\rho_{\tau\mu}$. The lower panels show the $\kappa_u$ is required
to be smaller than 1 due to the constraints of $B$-meson decays and
$R_b$. The upper panels show that there is a strong correlation
between $\rho_{\tau\mu}$ and $\kappa_\ell$, which is mainly due to
the constraints of $Br(\tau\to 3\mu)$ on the product
$|\rho_{\tau\mu}\times \kappa_\ell|$, and obviously affected
by the constraints of $Br(\tau\to\mu\gamma)$. For example, in the case of $m_A=126$ GeV,
 $|\rho_{\tau\mu}|$ is required to be smaller than 0.06 for
$\kappa_\ell=-10$.

In the case of $m_A=126$ GeV, there are two different regions where
the muon g-2 anomaly can be explained. (i) $\rho_{\tau\mu}=0$ and
$|\kappa_\ell|>100$: The Higgs LFV couplings are
absent due to $\rho_{\tau\mu}=0$, and the muon g-2 can be only corrected via the diagrams with
the Higgs LFC couplings. Without the contributions of top quark
loops, the contributions of the CP-even (CP-odd) Higgs to muon g-2
are negative (positive) at the two-loop level and positive
(negative) at one-loop level. As $m^2_f/m^2_\mu$ could easily
overcome the loop suppression factor $\alpha/\pi$, the two-loop
contributions may be larger than one-loop ones. Therefore, the muon g-2 can obtain the
positive contributions from $A$ loop and negative contributions from
$H$ loop. For the enough mass splitting of $H$ and $A$, the muon g-2
can be sizably enhanced by the diagrams with the large Higgs LFC
couplings. The corresponding $\kappa_u$ is required to be
smaller than 0.2 due to the constraints of the search for $gg\to A\to \tau\bar{\tau}$ at the LHC, see the pluses (red) with
$\rho_{\tau\mu}=0$ shown in the lower-left panel of Fig. \ref{cpvscp}. (ii) $0.04<|\rho_{\tau\mu}|<0.18$ and
$-9<\kappa_\ell<3$: The muon g-2 can be corrected by the diagrams with
the Higgs LFV interactions and the Higgs LFC interactions, and the contributions of the former dominate over those of the latter due
to the small $\mid\kappa_\ell\mid$. For the diagrams with the Higgs LFV couplings, the
muon g-2 obtains the positive contributions from $A$ loop and
negative contributions from $H$ loop due to
$\rho_{\mu\tau}=-\rho_{\tau\mu}$. For the enough mass splitting of
$H$ and $A$, the muon g-2 can be sizably enhanced by the diagrams
with the large Higgs LFV couplings, and slightly corrected by
those with the Higgs LFC couplings.

In the case of $m_H=126$ GeV, the contributions of the CP-even
Higgs dominate over those of the CP-odd Higgs due to $m_A > m_H$.
The muon g-2 obtains
 the negative contributions from the diagrams with the Higgs LFC couplings and
positive contributions from the diagrams with the LFV couplings due
to $\rho_{\mu\tau}=\rho_{\tau\mu}$. Therefore, a proper
$\rho_{\tau\mu}$ is required to explain the muon g-2 excess,
$0.04<|\rho_{\tau\mu}|<0.18$ and $-3<\kappa_\ell<8$ as shown in the
right panels of Fig. \ref{cpvscp}.

\begin{figure}[tb]
 \epsfig{file=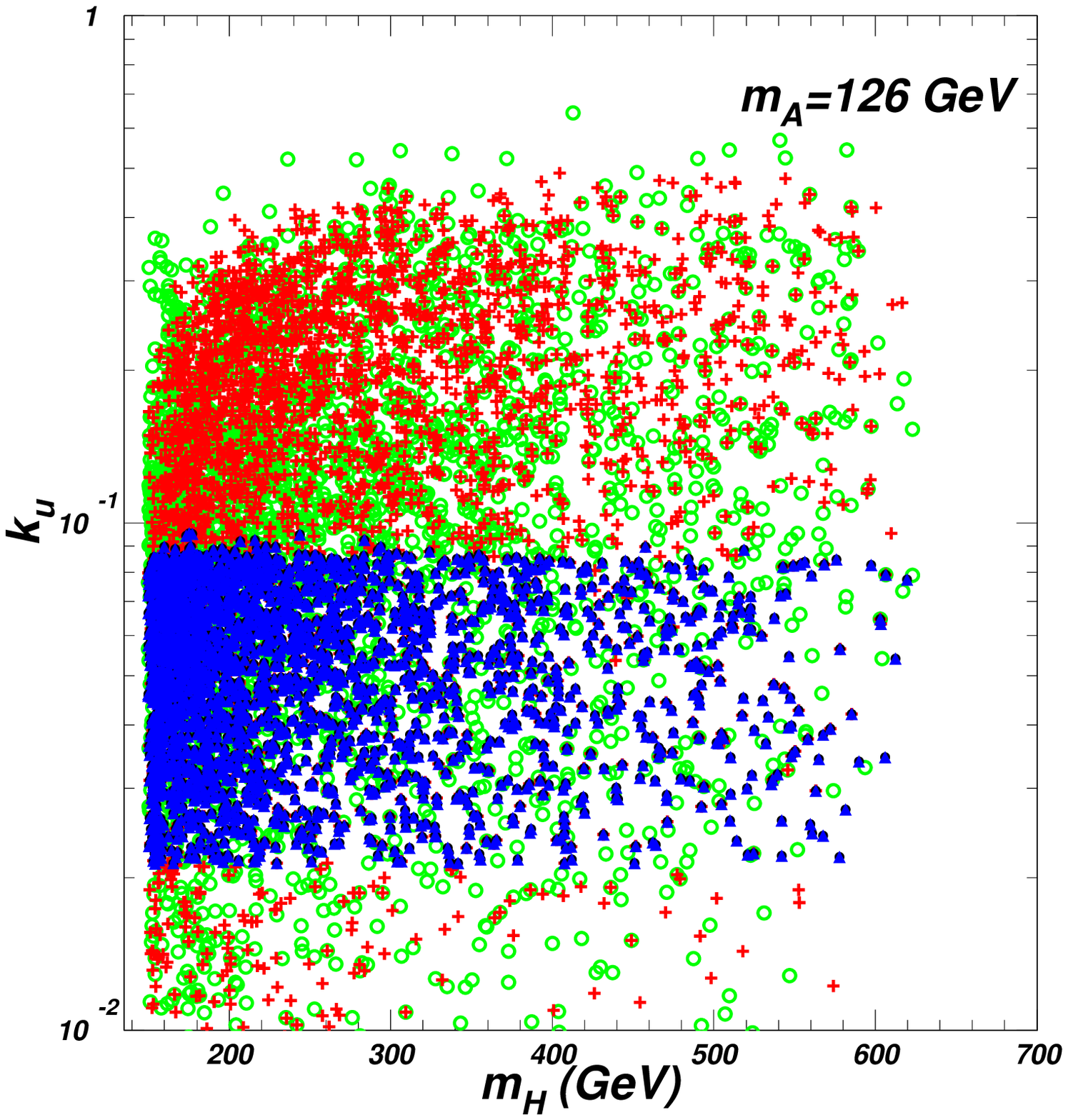,height=6.5cm}
  \epsfig{file=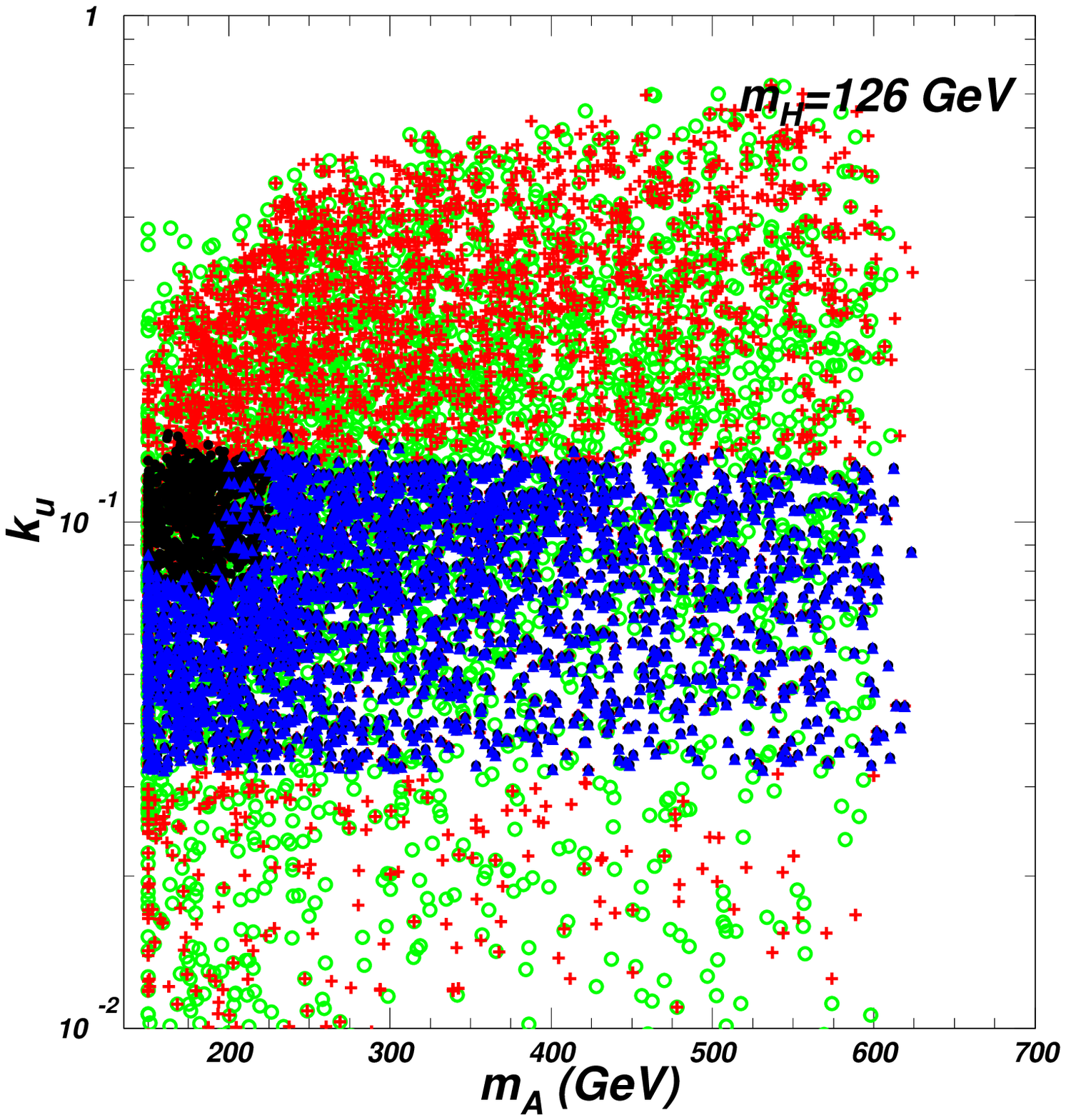,height=6.5cm}
\vspace{-0.5cm} \caption{Same as Fig. \ref{cpvscp}, but $\kappa_u$
versus $m_H$ and $\kappa_u$ versus $m_A$.} \label{csbigh}
\end{figure}

In the case of $m_A=126$ GeV, $0.02<\kappa_u<0.1$ is favored by
the $\mu\tau$ excess around 125 GeV and allowed by
the experimental constraints of the heavy Higgs decaying into $\mu\tau$. In the case of $m_H=126$ GeV,  $0.03<\kappa_u<0.15$
is favored by the $\mu\tau$ excess around 125 GeV, but some samples with a relatively large
$\kappa_u$ are excluded by the experimental constraints of the heavy Higgs decaying into $\mu\tau$. As
well known, the effective $ggA$ coupling is larger than the $ggH$ coupling for
the same Yukawa couplings and Higgs masses since the form factor of CP-odd
Higgs loop is larger than the CP-even Higgs loop. Thus, $\kappa_u$
in the case of $m_H=126$ GeV is required to be larger than that in the case of
$m_A=126$ GeV in order to obtain the correct $\mu\tau$ excess around 125 GeV.
$\sigma(pp\to A \to \mu\tau)$ in the case of $m_H=126$ GeV ($A$ as
the heavy Higgs) is much larger than $\sigma(pp\to H \to \mu\tau)$
in the case of $m_A=126$ GeV ($H$ as the heavy Higgs) due to the
enhancements of the large top Yukawa coupling and the form factor of
the CP-odd Higgs. Therefore, the experimental data of the heavy
Higgs decaying into $\mu\tau$ give more strong constraints on the case
of $m_H=126$ GeV than the case of $m_A=126$ GeV.

In Fig. \ref{csbigh}, we project the surviving samples on the plane
of $\kappa_u$ versus $m_A$ ($m_H$) in the case of $m_H=126$ GeV ($m_A=126$ GeV). The upper
bound of $\sigma(pp\to A/H \to \mu\tau)$ is taken from Ref.
\cite{xp1601.02616}, which is obtained by recasting results from the
original CMS $h\to\mu\tau$ analysis for the heavy Higgs in the range
of 125 GeV and 275 GeV. From the right panel, for the case of $m_H=126$ GeV
we find that the experimental data of the
heavy Higgs decaying into $\mu\tau$ can exclude most samples in the
ranges of $0.07<\kappa_u<0.15$ and $m_A<230$ GeV, which can explain
the excesses of muon g-2 and $\mu\tau$ around 125 GeV. For $m_A>230$
GeV, all the surviving samples which are consistent with the
$\mu\tau$ excess around 125 GeV are allowed by the experimental
constraints of the heavy Higgs decaying into $\mu\tau$. As discussed before, the left panel shows that 
all the surviving samples are allowed by the experimental constraints of the 
heavy Higgs decaying into $\mu\tau$ in the case of $m_A=126$ GeV.

Note that there is the $\kappa_\ell$ asymmetry in the regions of
$0.04<|\rho_{\tau\mu}|<0.18$, $-9<\kappa_\ell<3$ and
$0.02<\kappa_u<0.1$ for $m_A=126$ GeV where muon g-2 can be
explained. The main reason is from the constraints of
$\tau\to\mu\gamma$. In the above regions, the top quark can give 
sizable contributions to $\tau\to\mu\gamma$ via the $"A_{2L}"$ and
$"A_{2R}"$ terms as shown in Eq. (\ref{tamura2}), which have
destructive (constructive) interferences with the $"A_{1L0}"$ of Eq.
(\ref{tamura1}) and $"A_{1R0}"$ of Eq. (\ref{tamura1r})
induced by the one-loop contributions of $\tau$ for $\kappa_\ell<0$
($\kappa_\ell>0$). Therefore, $\mid\kappa_\ell\mid$ for
$\kappa_\ell<0$ is allowed to be much larger than that for
$\kappa_\ell>0$. Similar reason is for the $\kappa_\ell$ asymmetry in
the case of $m_H=126$ GeV but the destructive (constructive)
interferences for $\kappa_\ell>0$ ($\kappa_\ell<0$).

\begin{figure}[tb]
 \epsfig{file=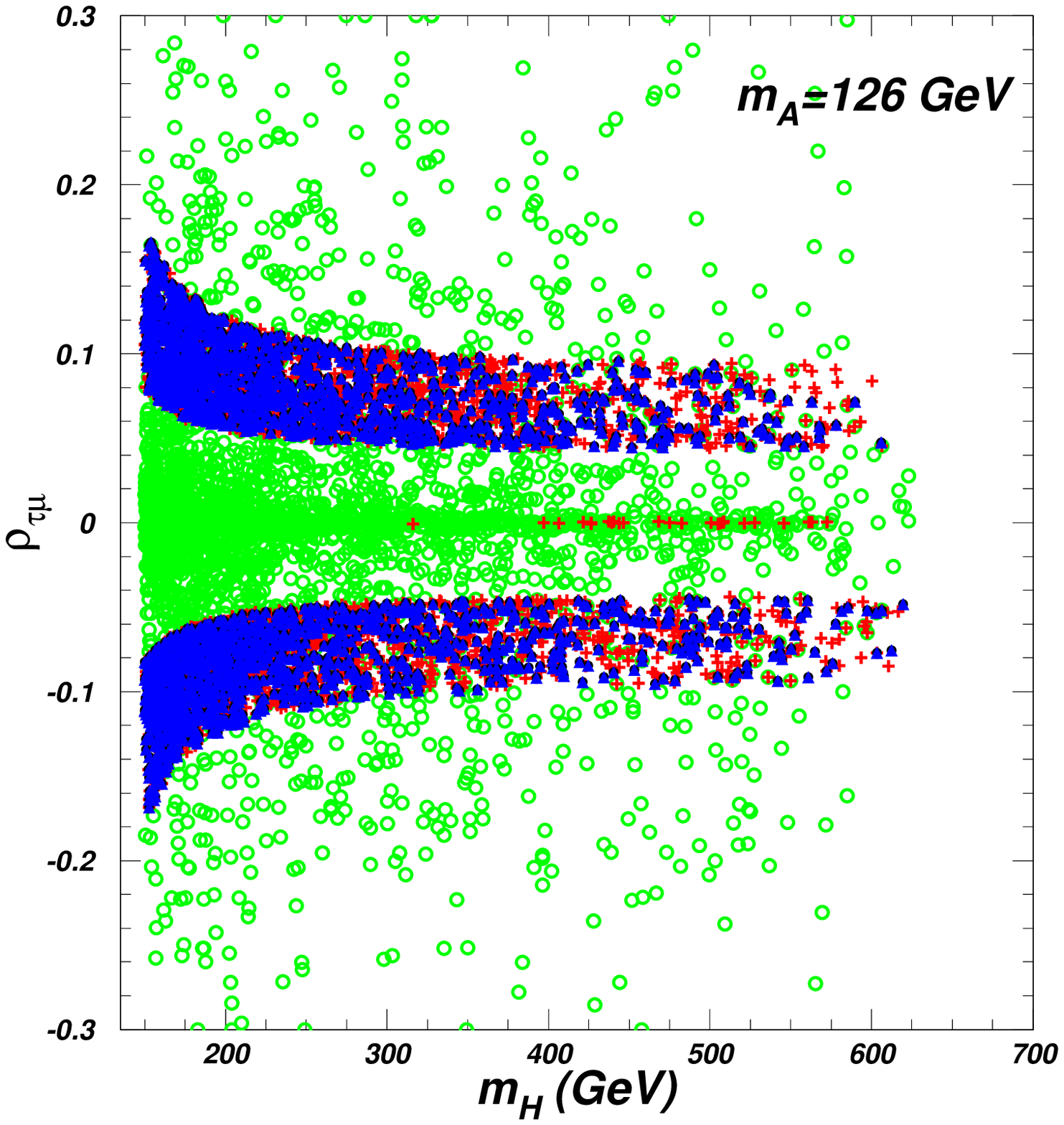,height=6.5cm}
  \epsfig{file=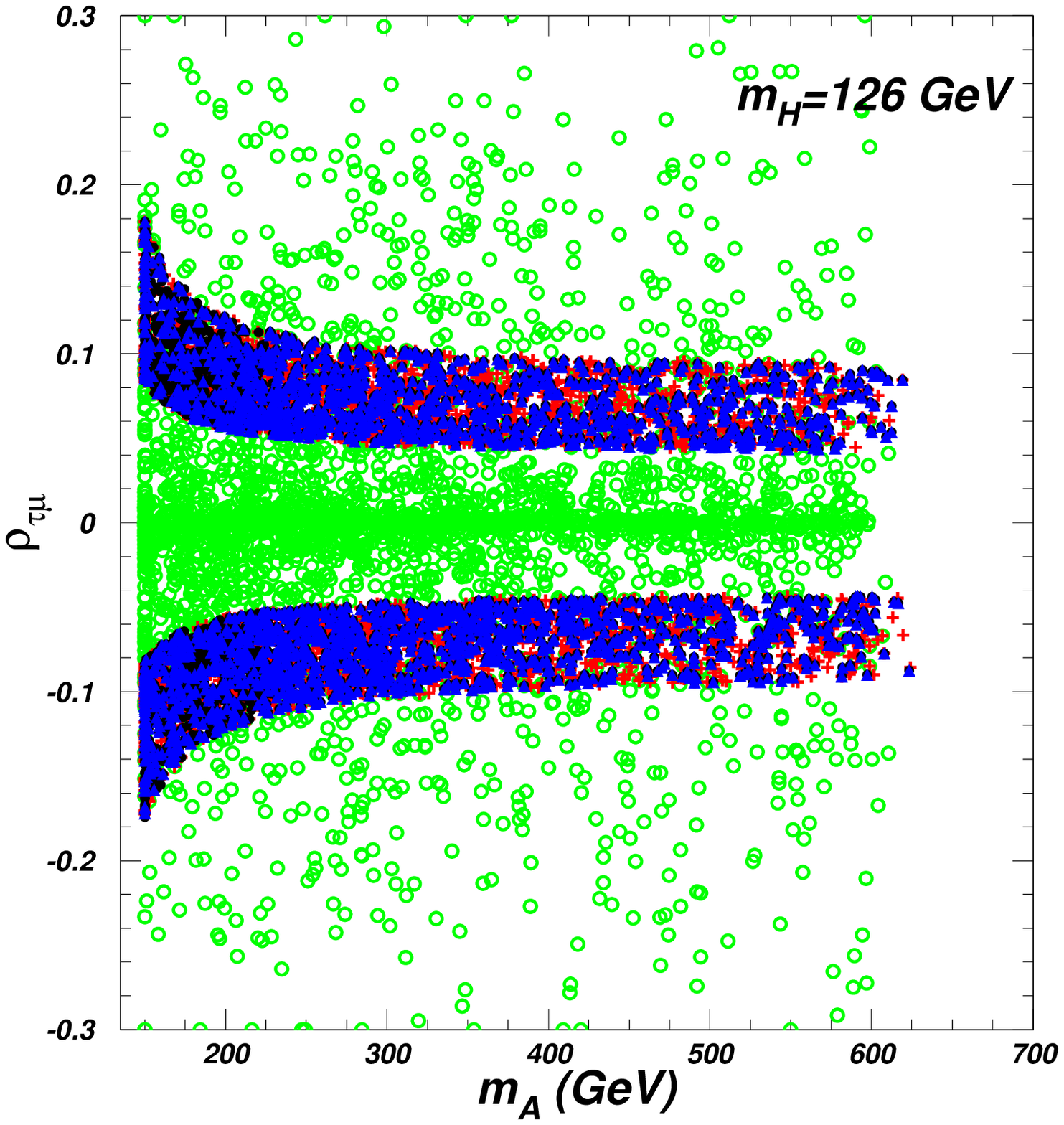,height=6.5cm}
\vspace{-0.5cm} \caption{Same as Fig. \ref{cpvscp}, but
$\rho_{\tau\mu}$ versus $m_H$ and $\rho_{\tau\mu}$ versus $m_A$.}
\label{cpmh}
\end{figure}

In Fig. \ref{cpmh}, we project the surviving samples on the planes
of $\rho_{\tau\mu}$ versus $m_H$ and $\rho_{\tau\mu}$ versus $m_A$
in the cases of $m_A=126$ GeV and $m_H=$ 126 GeV, respectively. We
find that $\rho_{\tau\mu}$ is sensitive to the mass of heavy Higgs,
and the absolute value decreases with increasing of the mass of
heavy Higgs in order to explain the muon g-2 anomaly and the $\mu\tau$ excess around
125 GeV. As we discussed above, there is an opposite sign between
the contributions of the $H$ loops and $A$ loops to the muon g-2.
Therefore, with the decreasing of the mass splitting of $H$ and $A$,
the cancelation between the contributions of $H$ and $A$ loops
becomes sizable so that a large absolute value of $\rho_{\mu\tau}$
is required to enhance the muon g-2.

\begin{figure}[tb]
 \epsfig{file=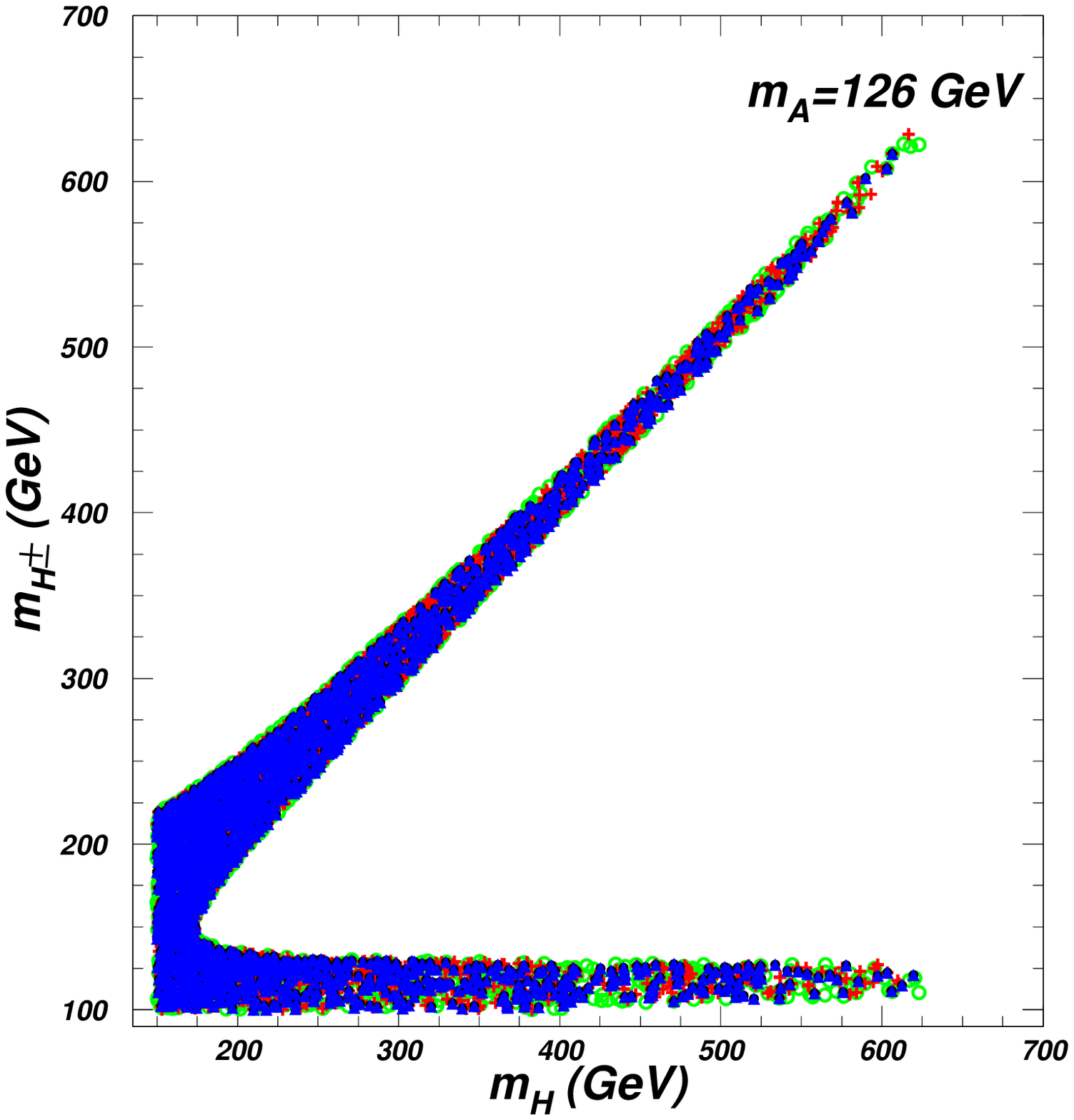,height=6.5cm}
 \epsfig{file=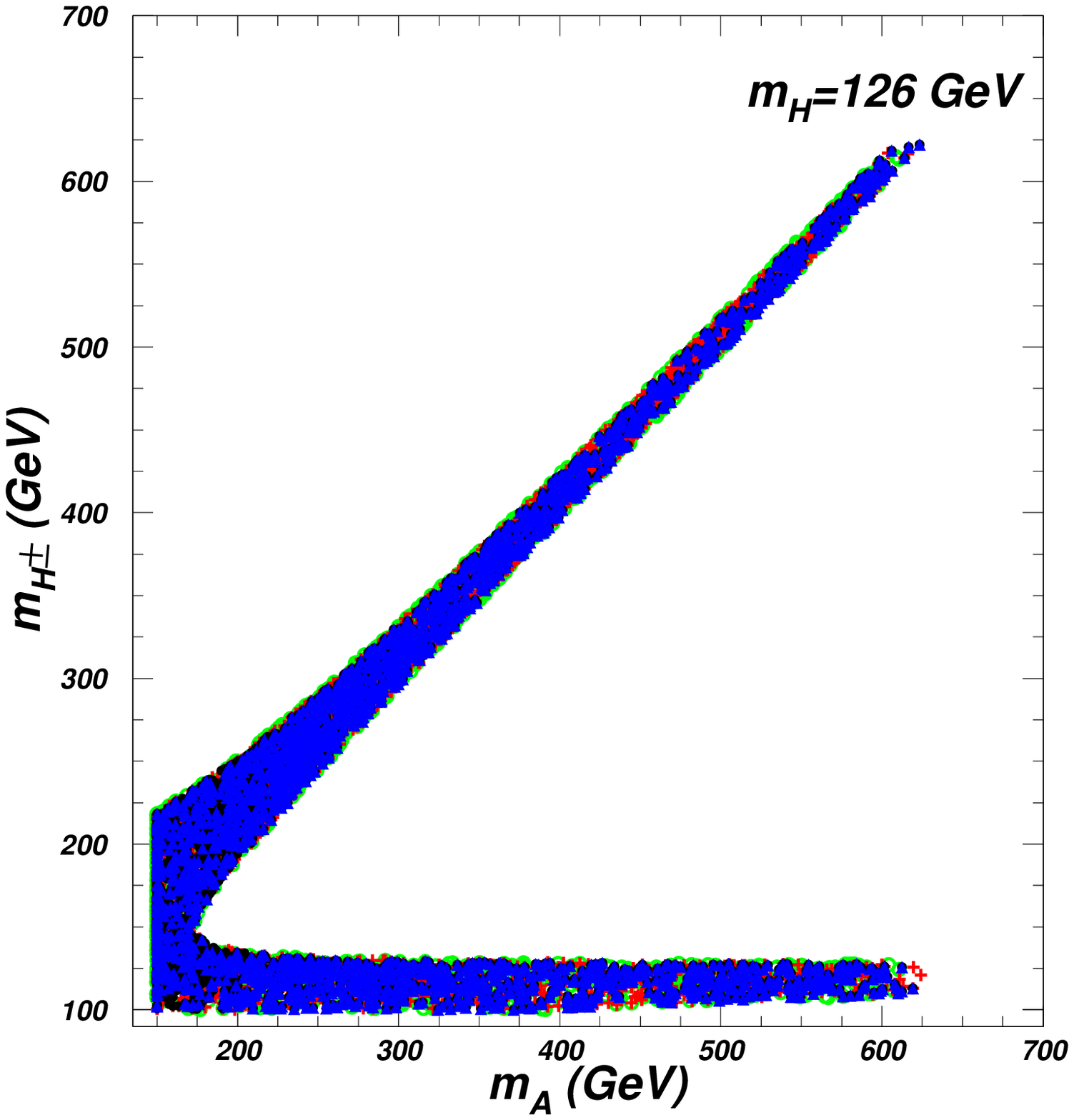,height=6.5cm}
\vspace{-0.5cm} \caption{Same as Fig. \ref{cpvscp}, but $m_{H^\pm}$
versus $m_H$ and $m_{H^\pm}$ versus $m_A$.} \label{hmass}
\end{figure}

In Fig. \ref{hmass}, we project the surviving samples on the planes
of $m_{H^{\pm}}$ versus $m_H$ and $m_{H^{\pm}}$ versus $m_A$ in the
cases of $m_A=126$ GeV and $m_H=$ 126 GeV, respectively. We find
that the mass splitting of $H^{\pm}$ and
 $H$ ($A$) decreases with increasing of $m_{H^{\pm}}$ in the case of
$m_A=126$ GeV ($m_H=$ 126 GeV), which is due to the constraints of
the oblique parameters and $\delta \rho$. However, for $m_{H^{\pm}}<130$
GeV, $m_H$ ($m_A$) is allowed to be as large as 625 GeV in the case
of $m_A=126$ GeV ($m_H=$ 126 GeV).

In this paper we focus on the exact alignment limit. If the
alignment limit is approximately realized, the $\mu\tau$ excess can
be from the SM-like Higgs ($h$) in addition to $H$ or $A$ around the
125 GeV. Therefore, the upper limits of $\kappa_u$ become more
stringent. When the $\mu\tau$ excess is mainly from $h$, the lower
limit of $\kappa_u$ will disappear since the $ht\bar{t}$ coupling
hardly changes with $\kappa_u$, and the $A(H)t\bar{t}$ coupling is
(nearly) proportional to $\kappa_u$. In addition, the upper limit of
$\rho_{\mu\tau}$ can become more strong for the proper deviation
from the alignment limit. For example, for
$\sin(\beta-\alpha)=0.996$, $Br(h\to\mu\tau)<1.62\%$ will give an
upper limit of $\mid\rho_{\mu\tau}\mid<0.0408$, which is much
smaller than that in the exact alignment limit. In the exact
alignment limit, the widths of $H\to hh,~WW^{(*)},~ZZ^{(*)}$ and
$A\to hZ$ are equal to zero, and increase with decreasing of
$\mid\sin(\beta-\alpha)\mid$. Therefore, the searches for $H\to
hh,~WW^{(*)},~ZZ^{(*)}$ and $A\to hZ$ can be used to probe the
deviation from the alignment limit. These signatures refer to the $H$ or $A$ whose mass is not near 125 GeV.
Otherwise, its signal would be indistinguishable from that coming from the SM-like light Higgs,
and even $H\to hh$ ($A\to hZ$) is absent for $H$ ($A$) near 125 GeV.
 Some similar studies have been
done in the singlet extension of the SM \cite{150102234}.

In the previous studies, the $\mu\tau$ excess is assumed to be from
the SM-like Higgs $h$. In this paper we discuss another interesting
scenario where the $\mu\tau$ excess is from either $H$ or $A$ near
the observed Higgs signal. There is no $AVV$ coupling due to the
CP-conserving. The $HVV$ coupling is absent and the $hVV$ coupling is the same as the SM value 
in the exact alignment limit.
Therefore, the two scenarios can be distinguished by observing the
$\mu\tau$ signal via the vector boson fusion production process at
the LHC with high integrated luminosity. In other words, if the
$\mu\tau$ signal excess is observed in the gluon fusion process and
not observed in the vector boson fusion process, the scenario in
this paper will be strongly favored. Even when $\sin(\beta-\alpha)$
deviates from the alignment limit sizably, the production rates of
$\mu\tau$ signal via the gluon fusion and vector boson fusion can
still have different correlations in the two different
scenarios.

\section{Conclusion}
In this paper we examine the muon g-2 anomaly and the $\mu\tau$
excess around 125 GeV in the exact alignment limit of 2HDM. In the scenario, the
SM-like Higgs couplings to the SM particles are
the same as the Higgs couplings in the SM at the tree level, and the tree-level LFV coupling $h\mu\tau$ is
absent. We assume the $\mu\tau$ signal excess observed by CMS to be
respectively from the $H$ and $A$, which almost degenerates with the
SM-like Higgs at the 125 GeV. After imposing various relevant
theoretical constraints and experimental constraints from precision
electroweak data, $B$-meson decays, $\tau$ decays and Higgs
searches, we obtain the following observations:

For the case of $m_A=126$ GeV, the muon g-2 anomaly can be explained
in two different regions: (i) $\rho_{\tau\mu}=0$ and
$|\kappa_\ell|>100$; (ii) $0.04<|\rho_{\tau\mu}|<0.18$
($|\rho_{\tau\mu}|$ is sensitive to $m_H$) and $-9<\kappa_\ell<3$.
Further, the $\mu\tau$ excess around 125 GeV can be explained in the
region (ii) with $0.02<\kappa_u<0.1$ where all the surviving samples
are allowed by the experimental constraints of the heavy Higgs decaying into $\mu\tau$.

For the case of $m_H=126$ GeV, the muon g-2 anomaly excludes the
region of $\rho_{\tau\mu}=0$, and can be only explained in the
region with a proper $\rho_{\tau\mu}$. The muon g-2 anomaly
and $\mu\tau$ excess favor $0.04<|\rho_{\tau\mu}|<0.18$ ($|\rho_{\tau\mu}|$ is sensitive to $m_A$),
$-3<\kappa_\ell<8$ and
$0.03<\kappa_u<0.15$. However, most samples in the ranges
of $0.07<\kappa_u<0.15$ and $m_A<230$ GeV are further excluded by the experimental constraints of the heavy
Higgs decaying into $\mu\tau$.

\section*{Acknowledgment}
This work has been supported by the National Natural Science
Foundation of China under grant Nos. 11575152, 11375248.

\appendix
\section{The measured values of the signal strengths of 125 GeV Higgs at the LHC and Tevatron.}
\begin{table}[thb!]
\caption{\small \label{aa} The measured values of the signal
strengths of $h\rightarrow \gamma \gamma$ at the LHC and Tevatron.
The composition $\epsilon^i_j$ of each production mode in each data
are given. }
\begin{ruledtabular}
\begin{tabular}{cccccccr}
Channel & Signal strength $\mu$ & $M_H$(GeV) & \multicolumn{4}{c}{Production mode}  &\\
        &      &            & ggF & VBF & VH & ttH \\
\hline
\multicolumn{8}{c}
{ATLAS (20.3fb$^{-1}$ at 8TeV)  \cite{atlas_aa_2014}}\\
\hline
$\mu_{ggH}$ & $1.32 \pm 0.38$ & 125.40 & 100\% & - & - & - & \\
$\mu_{VBF}$ & $0.8 \pm 0.7$ & 125.40 & - & 100\% & - & - &  \\
$\mu_{WH}$ & $1.0 \pm 1.6$ & 125.40 & - & - & 100\% & - &  \\
$\mu_{ZH}$ & $0.1^{+3.7}_{-0.1}$ & 125.40 & - & - & 100\% & - &  \\
$\mu_{ttH}$ & $1.6^{+2.7}_{-1.8}$ & 125.40 & - & - & - & 100\% &  \\
\hline
\multicolumn{7}{c}
{CMS (19.7fb$^{-1}$ at 8TeV)  \cite{cms_aa_2014} }\\
\hline
$\mu_{ggH}$ & $1.12^{+0.37}_{-0.32}$ & 124.70 & 100\% & - & - & - &  \\
$\mu_{VBF}$ & $1.58^{+0.77}_{-0.68}$ & 124.70 & - & 100\% & - & - &  \\
$\mu_{VH}$ & $-0.16^{+1.16}_{-0.79}$ & 124.70 & - & - & 100\% & - & \\
$\mu_{ttH}$ & $2.69^{+2.51}_{-1.81}$ & 124.70 & - & - & - & 100\% &  \\
\hline
\multicolumn{7}{c}
{Tevatron (10.0fb$^{-1}$ at 1.96TeV)  \cite{tev}}\\
\hline
Combined & $6.14^{+3.25}_{-3.19}$ & 125 & 78\% & 5\% & 17\% & - & \\

\end{tabular}
\end{ruledtabular}
\end{table}

\begin{table}[thb!]
\caption{\small \label{zz} The same as Table~\ref{aa} but for
$H\rightarrow Z Z^{(\ast)}$.}
\begin{ruledtabular}
\begin{tabular}{cccccccr}
Channel & Signal strength $\mu$ & $M_H$(GeV) & \multicolumn{4}{c}{Production mode}  & \\
        &       &            & ggF & VBF & VH & ttH \\
\hline
\multicolumn{8}{c}
{ATLAS (20.3fb$^{-1}$ at 8TeV)  \cite{atlas_zz_2014,kado_ichep}}\\
\hline
Inclusive & $1.66^{+0.45}_{-0.38}$ & 124.51 & 87.5\% & 7.1\% & 4.9\% & 0.5\% &
 \\
\hline
\multicolumn{7}{c}
{CMS (19.7fb$^{-1}$ at 8TeV) \cite{cms_zz_2014}}\\
\hline
Inclusive & $0.93^{+0.29}_{-0.25}$ & 125.6 & 87.5\% & 7.1\% & 4.9\% & 0.5\% &
 \\

\end{tabular}
\end{ruledtabular}
\end{table}

\begin{table}[thb!]
\caption{\small \label{ww} The same as Table~\ref{aa} but for
$H\rightarrow W W^{(\ast)}$.}
\begin{ruledtabular}
\begin{tabular}{cccccccr}
Channel & Signal strength $\mu$ & $M_H$(GeV) & \multicolumn{4}{c}{Production mode} &  \\
        &      &            & ggF & VBF & VH & ttH \\
\hline
\multicolumn{7}{c}
{ATLAS (20.7fb$^{-1}$ at 8TeV) \cite{mills_ichep} }\\
\hline
Inclusive & $0.99\pm 0.30$ & 125 & 87.5\% & 7.1\% & 4.9\% & 0.5\% &  \\
\hline
\multicolumn{7}{c}
{CMS (19.4fb$^{-1}$ at 8TeV) \cite{cms_ww_2014} }\\
\hline
0/1 jet & $0.74^{+0.22}_{-0.20}$ & 125.6 & 97\% & 3\% & - & - &  \\
VBF tag & $0.60^{+0.57}_{-0.46}$ & 125.6 & 17\% & 83\% & - & - &  \\
VH tag ($2l2\nu 2j$) & $0.39^{+1.97}_{-1.87}$ & 125.6 & - & - & 100\% & - &
 \\
WH tag ($3l3\nu$) & $0.56^{+1.27}_{-0.95}$ & 125.6 & - & - & 100\% & - &  \\
\hline
\multicolumn{7}{c}
{Tevatron (10.0fb$^{-1}$ at 1.96TeV) \cite{tev} }\\
\hline
Combined & $0.85^{+0.88}_{-0.81}$ & 125 & 78\% & 5\% & 17\% &  - & \\

\end{tabular}
\end{ruledtabular}
\end{table}
\begin{table}[thb!]
\caption{\small \label{bb} The same as Table~\ref{aa} but for
$H\rightarrow b\bar{b}$.}
\begin{ruledtabular}
\begin{tabular}{cccccccr}
Channel & Signal strength $\mu$ & $M_H$(GeV) & \multicolumn{4}{c}{Production mode}  & \\
        &       &            & ggF & VBF & VH & ttH \\
\hline
\multicolumn{8}{c}
{ATLAS (20.3${\rm fb}^{-1}$ at 8TeV) \cite{atlas_bb_2013,tth_ichep}}\\

\hline
VH tag & $0.2^{+0.7}_{-0.6}$ & 125.5 & - & - & 100\% & - & \\
ttH tag & $1.8^{+1.66}_{-1.57}$ & 125.4 & - & - & - & 100\% &  \\
\hline
\multicolumn{8}{c}
{CMS (18.9${\rm fb}^{-1}$ at 8TeV)
 \cite{cms_bb_2014}, (19.5${\rm fb}^{-1}$ at 8TeV) \cite{cms_bb_tth_2014}  }\\
\hline
VH tag & $1.0 \pm 0.5$ & 125 & - & - & 100\% & - &  \\
ttH tag & $0.67^{+1.35}_{-1.33}$ & 125 & - & - & - & 100\% & \\
\hline
\multicolumn{7}{c}
{Tevatron (10.0${\rm fb}^{-1}$ at 1.96TeV) \cite{tev_bb_2014} }\\
\hline
VH tag & $1.59^{+0.69}_{-0.72}$ & 125 & - & - & 100\% & - &  \\

\end{tabular}
\end{ruledtabular}
\end{table}
\begin{table}[thb!]
\caption{\small \label{t5}
The same as Table~\ref{aa} but for $H\rightarrow \tau \tau$.
The correlation for the $\tau\tau$ data of ATLAS is $\rho=-0.51$. }
\begin{ruledtabular}
\begin{tabular}{c c c cccc r}
Channel & \multicolumn{1}{c}{Signal strength $\mu$} & $M_H$(GeV) &
\multicolumn{4}{c}{Production mode} & \multicolumn{1}{c}{}
 \\
  &   && ggF & VBF & VH & ttH &  \\
\hline
\multicolumn{8}{c}
{ATLAS (20.3${\rm fb}^{-1}$ at 8TeV) \cite{atlas_h_tau_2013} }\\
\hline
$\mu(ggF)$  & $1.1^{+1.3}_{-1.0}$ & 125
 & 100\% & - & - & - & \\
$\mu(VBF+VH)$ &  $1.6^{+0.8}_{-0.7}$ & 125 & -
  & 59.6\% & 40.4\% & - &   \\
\hline
\multicolumn{8}{c}
{CMS (19.7${\rm fb}^{-1}$ at 8TeV) \cite{cms_tau_2014} }\\
\hline
0 jet &  $0.34 \pm 1.09$ & 125 &
$96.9\%$ & $1.0\%$ & $2.1\%$ & - &  \\
1 jet &  $1.07 \pm 0.46$ & 125 &
$75.7\%$ & $14.0\%$ & $10.3\%$ & - &   \\
VBF tag &  $0.94 \pm 0.41$ & 125 & $19.6\%$
 & $80.4\%$ & - & - &   \\
VH tag &  $-0.33 \pm 1.02$ & 125 & - & - & 100\% &
  - &   \\

\end{tabular}
\end{ruledtabular}
\end{table}

\end{document}